\documentclass[a4paper,11pt]{article}
\usepackage{microtype}
\usepackage{setspace}
\usepackage[left = 2cm, right=2cm, top=3cm, bottom=3cm]{geometry}
\usepackage[pdftex]{graphicx}
\usepackage{cite}
\usepackage{bm}
\usepackage{float}
\usepackage{amsmath,amssymb,latexsym}
\usepackage{color}
\usepackage{pdfpages}
\usepackage[T1]{fontenc}
\usepackage{wasysym}
\usepackage{siunitx}
\usepackage{setspace}
\usepackage{relsize}
\usepackage{tikz}
\usepackage{lineno, blindtext}
\usepackage{authblk}
\usepackage[symbol]{footmisc}
\usepackage{lineno}
\usepackage{chemformula}
\usepackage{graphicx}
\usepackage{tabularx}
\usepackage{makecell}
\usepackage[pdfencoding=auto, psdextra]{hyperref}

\usepackage{silence}
\begin{document}
	\renewcommand{\figurename}{FIGURE}
	\title{\color{blue}\textbf{Tuning Thermal Memory and Dissipation in Colloidal Microgel Glasses via Particle Stiffness}}
	\author[1]{Sonali Vasant Kawale}
	\affil[1]{\textit{Soft Condensed Matter Group, Raman Research Institute, C. V. Raman Avenue, Sadashivanagar, Bangalore 560 080, INDIA}}
	\author[1,*]{Ranjini Bandyopadhyay}
	\footnotetext[1]{Corresponding Author: Ranjini Bandyopadhyay; Email: ranjini@rri.res.in}
	\maketitle
	\begin{abstract}
{The Kovacs effect, characterized by nonlinear, path-dependent structural recovery,  is a hallmark of glassy systems. In soft colloidal microgel glasses, particle stiffness governs deformation, swelling, rearrangement, and response to external stimuli, and thereby dictates the path-dependent structural recovery of these jammed non-equilibrium systems. To investigate this influence, microgels of different stiffnesses were synthesized by adjusting the concentration of the polar crosslinker, $N, N^{\prime}$-methylenebisacrylamide (MBA). The relaxation of the storage and loss moduli was measured using oscillatory rheology under diverse temperature ramp rates, and cryo-FESEM imaging provided direct visualization of the underlying particle packing. While the rapid adaptability of soft particles and the kinetic arrest of stiff ones suppresses Kovacs asymmetry, intermediate-stiffness particles exhibit moderate thermoresponsivity, traversing distinct energy landscapes on heating and cooling to maximize asymmetry and dissipation. Cryo-FESEM micrographs, displaying gel-like networks in soft particle suspensions, dense packing in stiff particle suspensions, and disordered, compact structures in intermediate-stiffness systems, corroborate these findings. Our results demonstrate that intermediate-stiffness PNIPAM microgels optimize structural recovery, rendering them promising candidates for forming colloidal glasses, and establish Kovacs asymmetry as a robust probe of non-equilibrium dynamics in glassy systems.}     

        \end{abstract}                   
	\noindent
	\definecolor{black}{rgb}{0.0, 0.0, 0.0}
	\definecolor{red(ryb)}{rgb}{1.0, 0.15, 0.07}
	\definecolor{darkred}{rgb}{0.55, 0.0, 0.0}
	\definecolor{blue(ryb)}{rgb}{0.01, 0.2, 1.0}
	\definecolor{darkcyan}{rgb}{0.0, 0.55, 0.55}
	\definecolor{navyblue}{rgb}{0.0, 0.0, 0.5}
	\definecolor{olivedrab(web)(olivedrab3)}{rgb}{0.42, 0.56, 0.14}
	\definecolor{darkraspberry}{rgb}{0.53, 0.15, 0.34}
	\definecolor{magenta}{rgb}{1.0, 0.0, 1.0}
	
	\newcommand{\blsquare}{\textcolor{black}{\small$\blacksquare$}}
	\newcommand{\hlsquare}{\textcolor{darkred}{\small$\square$}}
	\newcommand{\redtraingle}{\textcolor{magenta}{\small$\triangle$}}
	\newcommand{\oolive}{\textcolor{olivedrab(web)(olivedrab3)}{\large$\circ$}}
	\newcommand{\purpletraingle}{\textcolor{darkraspberry}{\small$\triangledown$}}
	
	\newcommand{\bltriangle}{\textcolor{black}{\small$\triangleup$}}
	\newcommand{\rcircle}{\textcolor{red(ryb)}{\large$\bullet$}}
	\newcommand{\rtraingle}{\textcolor{red(ryb)}{\small$\triangledown$}}
	\newcommand{\wine}{\textcolor{darkred}{\large$\bullet$}}
	\newcommand{\cyan}{\textcolor{darkcyan}{\large$\bullet$}}
	\newcommand{\owine}{\textcolor{darkred}{\large$\circ$}}
	\newcommand{\redcircle}{{\large$\circ$}}
	\newcommand{\ocyan}{\textcolor{darkcyan}{\large$\circ$}}
	\newcommand{\blue}{\textcolor{blue(ryb)}{\large$\bullet$}}
	\newcommand{\blbullet}{\textcolor{navyblue}{\large$\bullet$}}
	\newcommand{\olbullet}{\textcolor{olivedrab(web)(olivedrab3)}{\large$\bullet$}}
	\newcommand{\hollowblue}{\textcolor{blue(ryb)}{\large$\circ$}}
	\newcommand{\black}{\textcolor{black}{\large$\bullet$}}
	\newcommand{\hollowblack}{\textcolor{black}{\large$\circ$}}
	\newcommand{\bltria}{\textcolor{black}{\small$\triangle$}}
	\newcommand{\rtrai}{\textcolor{red(ryb)}{\large$\triangledown$}}	
	\section{Introduction:}
 Microstructurally arrested materials, spanning molecular glasses to soft glassy systems, retain a pronounced memory of their thermal or mechanical (shear) history, which determines their subsequent recovery toward thermodynamic equilibrium \cite{vermeij1995through, tool1946viscosity, bertin2003kovacs,mckenna1999tau, plazek1980temperature, larson1999structure, angell2000relaxation, van2017fragility, yunker2014physics}. Understanding and controlling this thermal and/or mechanical history-dependence yields valuable insights into microstructural dynamics and is central to the processing and stability of colloidal formulations. Microstructural recovery in glassy systems demonstrates pronounced nonlinearity and is strongly influenced by thermal history, as highlighted by the Kovacs effect \cite{10.1007/BFb0050366, mckenna1999tau, mckenna201750th, larson1999structure, scalliet2019rejuvenation}. In molecular glasses, memory originates from structural characteristics dictated by chemical composition \cite{angell1995formation, debenedetti2001supercooled}. In contrast, memory in colloidal glasses can be engineered through particle design. Moreover, colloidal suspensions facilitate the investigation of glassy dynamics at experimentally accessible time and length scales, while closely replicating the structural arrest and relaxation observed in molecular systems \cite{pusey1986phase, mattsson2009soft, di2011signatures, vlassopoulos2014tunable, yunker2009irreversible, yunker2014physics, suman2021rheological}.
 
Microgels are colloidal particles defined by their intrinsic softness and deformability. At elevated concentrations, the constituent particles shrink, deform, and interpenetrate, with particle softness emerging as a critical factor governing their collective behavior \cite{di2011signatures, di2014dynamics, peng2016physical, kawale2026thermal}. Poly(N-isopropylacrylamide) (PNIPAM) microgel is a classic thermoresponsive synthetic colloidal system that exhibits reversible swelling-deswelling transitions in aqueous media in response to temperature change. When suspended in water, PNIPAM particles display a volume phase transition temperature (VPTT) of approximately 34$^{\circ}$C, which enables precise control over the effective volume fraction of their suspensions \cite{dong1991novel, dufresne2004preparation, romeo2010temperature, vlassopoulos2014tunable, franco2021glass, franco2025soft, yunker2014physics, town2019understanding, scotti2020flow, scotti2022softness, scalliet2019rejuvenation}. Furthermore, the stiffness of these materials can be independently controlled by varying the concentration of the polar crosslinker ($N,N^{\prime}$-methylenebisacrylamide, MBA) during synthesis \cite{mcphee1993poly, senff2000influence, misra2020influence, scotti2022softness}.

 Since the stiffness of individual particles governs their ability to deform and interpenetrate, it fundamentally governs particle-scale dynamics in dense packings. Romeo \textit{et al.} \cite{romeo2010temperature} demonstrated that dense PNIPAM suspensions constituted by particles of a uniform stiffness form a repulsive glass at temperatures below the VPTT, liquefy at the VPTT, and form a gel structure \textit{via} hydrophobic aggregation upon further heating. 
Using rheology and scanning electron microscopy, Misra \textit{et al.} \cite{misra2024effect} demonstrated that the phase behavior of PNIPAM microgels is governed by particle stiffness. They reported that suspensions of softer particles exhibit a gel–viscoelastic liquid–gel transition across the VPTT, whereas suspensions of stiffer particles undergo a glass–glass transition under the same conditions. Another study demonstrated that highly crosslinked particles exhibit strong interparticle interactions and are characterized by high elasticity, while weak crosslinking results in delayed yielding due to reduced surface chain interpenetration \cite{senff2000influence}. Mattsson \textit{et al.} \cite{mattsson2009soft} showed that the stiffness of PNIPAM-PAAc microgel particles governs the fragility of dense suspensions. Softer particles produce less fragile glassy states, while stiffer particles, which resemble hard spheres, lead to more fragile glassy states. The critical role of particle stiffness in determining the linear and non-linear viscoelastic properties of PNIPAM microgel suspensions was recently confirmed by Vialetto \textit{et al.} \cite{vialetto2024effect}. While these results clearly indicate that suspension phase behavior can be precisely regulated by adjusting the stiffness of PNIPAM particles, its influence on thermal memory and path-dependent structural recovery is less understood.

In out-of-equilibrium systems, the relaxation toward equilibrium is highly complex and exhibits pronounced path-dependent asymmetries in heating and cooling experiments. Pioneered by Kovacs using temperature-jump protocols on polymer glasses~\cite{10.1007/BFb0050366, tool1946viscosity, struik1978physical, larson1999structure, scalliet2019rejuvenation, bertin2003kovacs, yunker2009irreversible}, this asymmetry arises because structural recovery depends on both temperature and instantaneous structure 
\cite{mckenna1999tau, lira2021fundamental, mckenna201750th, kaushal2016analyzing, agarwal2020signatures, kawale2026thermal, sumannorm}. 
Such history-dependent relaxation is a universal hallmark of disordered and arrested matter, extending from core–shell microgel dispersions~\cite{di2014dynamics, peng2016physical} and filled elastomers~\cite{robin2022glass} to aging nanocolloidal gels \cite{chen2023memory}. Although glassy dynamics in soft colloids have been extensively studied, the extent to which single-particle stiffness encodes thermal history and governs Kovacs asymmetry remains an open question. 
 
To elucidate the role of particle stiffness in thermal memory, PNIPAM particles of different stiffnesses were synthesized by adjusting the concentration of a polar crosslinker. Dense PNIPAM suspensions prepared at a fixed effective volume fraction were subjected to heating and cooling ramps at different rates below the VPTT while monitoring the storage (G$^\prime$) and loss (G$^{\prime\prime}$) moduli. Our data revealed a pronounced, non-monotonic asymmetry in G$^\prime$ with increasing particle stiffness, with maximum asymmetry observed in intermediate-stiffness particle suspensions. Interestingly, G$^{\prime\prime}$ displayed peaks at the initiation and termination of the temperature ramp, signifying particle rearrangement events whose integrated areas exhibited a similar non-monotonic trend with increasing particle stiffness. Soft particles, owing to their high adaptability, readily accommodate thermal changes, yielding minimally path-dependent states, whereas stiff particles undergo negligible volume change and kinetic arrest. Consequently, both soft and stiff particle suspensions suppress asymmetry and dissipation. In contrast, intermediate-stiffness microgels traverse different energy landscapes during heating and cooling, resulting in maximal asymmetry and dissipation. Cryo-FESEM imaging corroborates these findings, showing loose gel-like networks for soft particles, dense packings for stiff ones, and disordered, compact structures that retain accessible free volume for intermediate-stiffness particles. We conclude that microgel particle stiffness dictates nonlinear structural recovery, with intermediate-stiffness particle suspensions emerging as optimal candidates for investigating memory effects in colloidal glasses.

\maketitle
\section{Material and Methods}
\subsection{Synthesis of PNIPAM microgel particles of different stiffnesses}
All chemicals used for particle synthesis were procured from Sigma-Aldrich and used as supplied, without further purification. 7.0 g of $N$-isopropylacrylamide (NIPAM, 99\%) monomer, the appropriate amount of a polar crosslinker $N, N'$-methylenebisacrylamide (MBA, 99.5\%), and 0.03 g sodium dodecyl sulfate (SDS) were dissolved in 470 ml Milli-Q water (Millipore Corp.) in a three-necked round-bottom flask equipped with a reflux condenser, magnetic stirrer, platinum sensor, and nitrogen inlet/outlet. The mixture was stirred at 600 rpm and purged with nitrogen (N$_2$) gas for 30 min to remove dissolved oxygen. Polymerization, initiated by adding 0.28 g potassium persulfate (KPS, 99.9\%) dissolved in 30 ml Milli-Q water at 70${^\circ}$C, was allowed to proceed for 4 hours with continuous stirring. After completion, the suspension was cooled to room temperature. PNIPAM particles of different stiffnesses~\cite{mcphee1993poly, misra2024effect} were synthesized by adjusting the crosslinker (MBA) content between 1 and 20\% of the NIPAM monomer mass. Increasing crosslinker-monomer concentration ratio yields stiffer microgel particles. The suspension was purified by four cycles of centrifugation at 15,000 to 50,000 rpm for 60 minutes each, followed by re-dispersion to eliminate SDS, residual monomers, oligomers, and other impurities. The supernatant was discarded, and the product was dried by evaporation of water, then ground into a fine powder using a mortar and pestle. In this study, particle stiffness is expressed as the percentage of MBA relative to NIPAM. Table~\ref{table:Synthesis} summarizes the synthesis protocols and characteristics of particles of all stiffnesses.
\begin{table*}
			\begin{center}
				\resizebox{\textwidth}{!}{%
                  \begin{tabular}{|c|c|c|c|c|c|c|c|c|}
					\hline

		$NIPAM$ &  $KPS$ & $SDS$ & $MBA/NIPAM$ & Stiffness Category & $\alpha$ & $c_p$ & $\phi_{eff}$ & $c$ \\
				(g) & (g) & (g) & $($\%$)$ & & $(d_{fullswell}/d_{fullshrunk})$   & $(wt.\%)$ & & $(wt.\%)$ \\
				\hline
				7 & $0.28$ & $0.03$ & $1.0$ & Soft & $2.43\pm 0.070$ & $2.42$ & $1.6$ & $3.87$ \\
				\hline
				7 & $0.28$ & $0.03$ &  $5.0$ & Soft & $1.85\pm 0.064$ & $4.95$ & $1.6$ & $7.92$ \\
				\hline
				7 & $0.28$ & $0.03$ & $7.5$ & Intermediate & $1.66\pm 0.051$ & $7.95$ & $1.6$ & $12.72$ \\
				\hline
				7  & $0.28$ & $0.03$ & $10.0$ & Intermediate & $1.59\pm 0.021$ & $10.37$ & $1.6$ & $16.59$ \\
				\hline
				7  & $0.28$ & $0.03$ & $12.5$ & Intermediate & $1.48\pm 0.043$ & $12.20$ & $1.6$ & $19.52$ \\
				\hline
				7  & $0.28$ & $0.03$ & $15.0$ & Stiff & $1.40\pm 0.031$ & $15.33$ & $1.6$ & $24.53$ \\
				\hline
				7  & $0.28$ & $0.03$ & $20.0$ & Stiff & $1.30\pm 0.029$ & $20.22$ & $1.6$ & $32.35$ \\
				\hline
				\end{tabular}%
                 }
				\caption{Ingredients used to synthesize PNIPAM particles and characterization of PNIPAM swelling properties/stiffnesses. $c_p$, $\phi_{eff}$, $c$ are, respectively, polymer mass concentration inside each PNIPAM particle, effective volume fraction, and PNIPAM particle concentration in aqueous suspension. $\phi_{eff}$ and $c$ were obtained from rheological measurements, and are discussed in Section 2 of the supporting information.}
				\label{table:Synthesis}
			\end{center}
		\end{table*}

\subsection{Dynamic light scattering}
\begin{figure*}
    \centering
    \includegraphics[width=\linewidth]{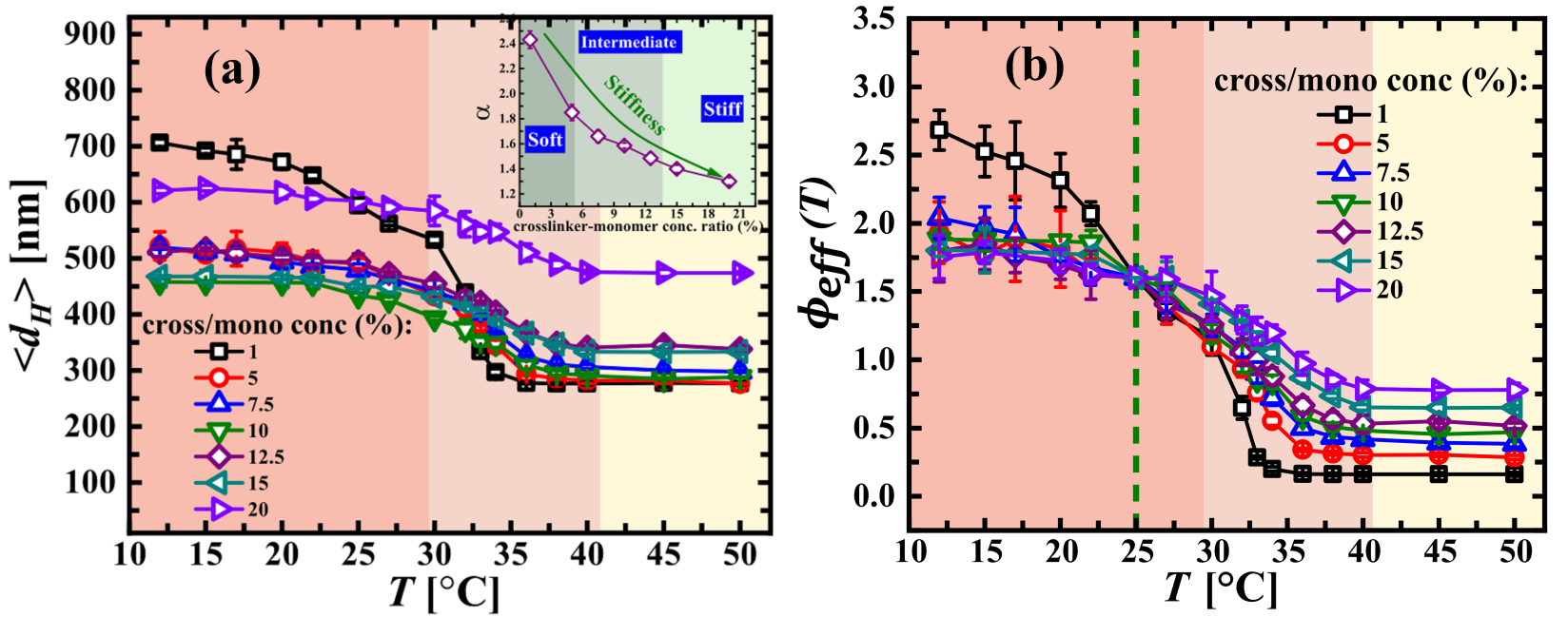}
    \caption{(a) Temperature-dependent average hydrodynamic diameters, $\langle d_H \rangle$, of PNIPAM microgels constituted by particles of different stiffnesses (crosslinker-monomer concentration ratios labeled as cross/mono conc in figure legend) in aqueous suspension. The inset shows the maximum swelling ratios, $\alpha=d_{20^{\circ}C}/d_{50^{\circ}C}$, of PNIPAM particles \textit{vs.} PNIPAM particle stiffness. The green arrow in the inset points toward increasing particle stiffness, with the soft, intermediate, and stiff regimes represented by different shades of green. (b) Temperature-dependent effective volume fractions, $\phi_{eff}$(\textit{T}), of dense aqueous suspensions (prepared at $\phi_{eff} =$ 1.6 at 25$^{\circ}$C, indicated by the vertical green dashed line) constituted by PNIPAM particles of different stiffnesses.} 
    \label{fig:schematicphieffective}
\end{figure*}

 Dynamic light scattering (DLS) measurements \cite{berne2000dynamic} were conducted to determine the average hydrodynamic diameter, $<$$d_{H}$$>$, of thermoresponsive PNIPAM particles over a temperature range of $12^{\circ}$C to $50^{\circ}$C. A detailed experimental procedure is provided in Section 1 of the supporting information. Very dilute suspensions (\(\phi \approx 10^{-3}\)) were prepared by dispersing PNIPAM particles of different crosslinker-monomer concentration ratios, and therefore of different stiffnesses, in Milli-Q water. Figure~\ref{fig:schematicphieffective}(a) shows the dependence of $<$$d_{H}$$>$ on temperature for the particles of different stiffnesses studied in this work. The stiffness of PNIPAM particles increased monotonically with concentration of the crosslinker MBA(\%) relative to the monomer NIPAM concentration incorporated during synthesis. The DLS data were further evaluated to calculate the maximum swelling ratio, $\alpha = d_{20^{\circ}C}/d_{50^{\circ}C}$, where $d_{20^{\circ}C}$ and $d_{50^{\circ}C}$ represent the mean hydrodynamic diameters of fully swollen and collapsed PNIPAM particles at $20^{\circ}$C and $50^{\circ}$C, respectively. The stiffness is inversely related to $\alpha$ (see inset of Figure~\ref{fig:schematicphieffective}(a) and Table~\ref{table:Synthesis}). Figure~S1(b) of the supporting information displays the temperature-dependent swelling ratios, $\alpha_{T}=d_{T^{\circ}C}/d_{50^{\circ}C}$, for particles of different stiffnesses, where $d_{T}$ denotes the mean hydrodynamic diameter at temperature $T$. The markedly larger swelling and deswelling of softer particles across the VPTT is attributed to their lower crosslink density, which reduces elastic resistance to osmotic pressure and facilitates water intake and expulsion \cite{sierra2012phase}.
 
\subsection{Preparation of dense aqueous PNIPAM suspensions}
 Dense suspensions of PNIPAM microgel particles of different stiffnesses were prepared by dispersing PNIPAM powder in Milli-Q water. The mixture was stirred for 24 hours and then sonicated for 15 minutes. The resulting suspensions were stored at  4$^{\circ}$C for further use. Due to the deformability and compressibility of PNIPAM particles, the conventional volume fraction, $\phi$, does not accurately represent particle packing in the dense suspension. Therefore, the effective volume fraction $\phi_{eff}$, calculated as described in Section~2 of the supporting information, is used throughout this study. For the first set of experiments, the concentrations of ingredients were adjusted appropriately to achieve $\phi_{eff} = 1.6$ at $T = 25^{\circ}$C. Since stiffer particles are less deformable and compressible than softer particles, this preparation protocol required us to use higher concentrations of ingredients to prepare stiff-particle suspensions (Table~\ref{table:Synthesis}). The temperature dependence of $\phi_{eff}(T)$ was determined using the following relation \cite{carrier2009nonlinear}:
\begin{equation}
{\phi_{eff}(T) = \phi_{eff}(25^\circ \text{C}) \times \left( \frac{d_H(T)}{d_H(25^\circ \text{C})} \right)^3 } 
  \label{eqn}
\end{equation}
where $\phi_{eff}(T)$ and $\phi_{eff}(25^\circ \text{C})$ are the effective volume fractions, while $d_H(T)$ and $d_H(25^\circ \text{C})$ are the average hydrodynamic diameters, at temperatures $T$ and $25^\circ \text{C}$, respectively. Figure~\ref{fig:schematicphieffective}(b) shows the temperature-dependent effective volume fractions, $\phi_{eff}(T)$, of PNIPAM suspensions constituted by particles of different stiffnesses. Below the VPTT, the microgels are fully swollen and form densely packed structures. As the temperature approaches the VPTT, water is expelled from PNIPAM particles regardless of their stiffness, leading to a reduction in the effective volume fraction and subsequent suspension unjamming. Soft particles unjam more rapidly than stiff particles (Figure~S1(b)). At temperatures well above the VPTT, microgels remain collapsed and aggregate into gels, primarily due to enhanced hydrophobic interactions between particles \cite{romeo2010temperature,misra2024effect}. 

Due to differences in deswelling behavior with changing particle stiffness, maintaining a constant effective volume fraction ($\phi_{eff} = 1.6$ at $25^\circ\text{C}$) leads to substantial variations in particle concentration across suspensions, especially at high temperatures lying above the VPTT. To isolate the effects of these concentration differences, control suspensions of representative soft (1\% MBA/NIPAM) and stiff (15\% MBA/NIPAM) microgels were also prepared, with their particle number densities matched to that of the 10\% MBA/NIPAM reference suspension used earlier ($\phi_{eff} = 1.6$ at 25$^{\circ}$). The concentrations of these suspensions were determined by measuring the average hydrodynamic diameters of the particles in each sample by dynamic light scattering (DLS). These samples were studied to verify the earlier results, and their effective volume fractions were measured to be $\phi_{eff} = 3.95$ for the soft microgels and $\phi_{eff} = 1.55$ for the stiff microgels at $25^\circ\text{C}$.

\subsection{Rheology}
All rheological measurements \cite{macosko1994rheology} were carried out with a stress-controlled Anton Paar MCR 702 rheometer. Flow curves for dilute PNIPAM suspensions were measured using a double gap geometry (gap width 1.886 mm, effective length 40 mm, sample volume 3.8 ml). This data was analyzed to estimate the effective volume fractions of the dense PNIPAM suspensions (see Section 2 of the supporting information). For dense suspensions, oscillatory measurements were performed using a cone and plate geometry (cone radius \(r_{C}\) = 12.491 mm, cone angle 0.979$^{\circ}$, gap d = 0.048 mm, sample volume of 0.07 ml). The sample temperature was controlled with a Peltier unit for the cone-plate geometry (accessible range 0$^{\circ}$C to 180$^{\circ}$C) and a Viscotherm VT2 water circulation unit for the double-gap geometry (range 5$^{\circ}$C to 80$^{\circ}$C). Evaporative water loss was minimized in both geometries by using a silicone oil solvent trap (viscosity 5 cSt). The viscoelastic responses of dense PNIPAM suspensions constituted by particles of different stiffnesses were examined using oscillatory rheological measurements under temperature ramps applied at different rates. Prior to each measurement, a rejuvenation step was applied to eliminate the loading history. All oscillatory measurements were conducted using a peak-to-peak strain amplitude, $\gamma = 0.5\%$, and an angular frequency of 10 rad/s.

\begin{figure*}
    \centering
    \includegraphics[width=0.5\linewidth]{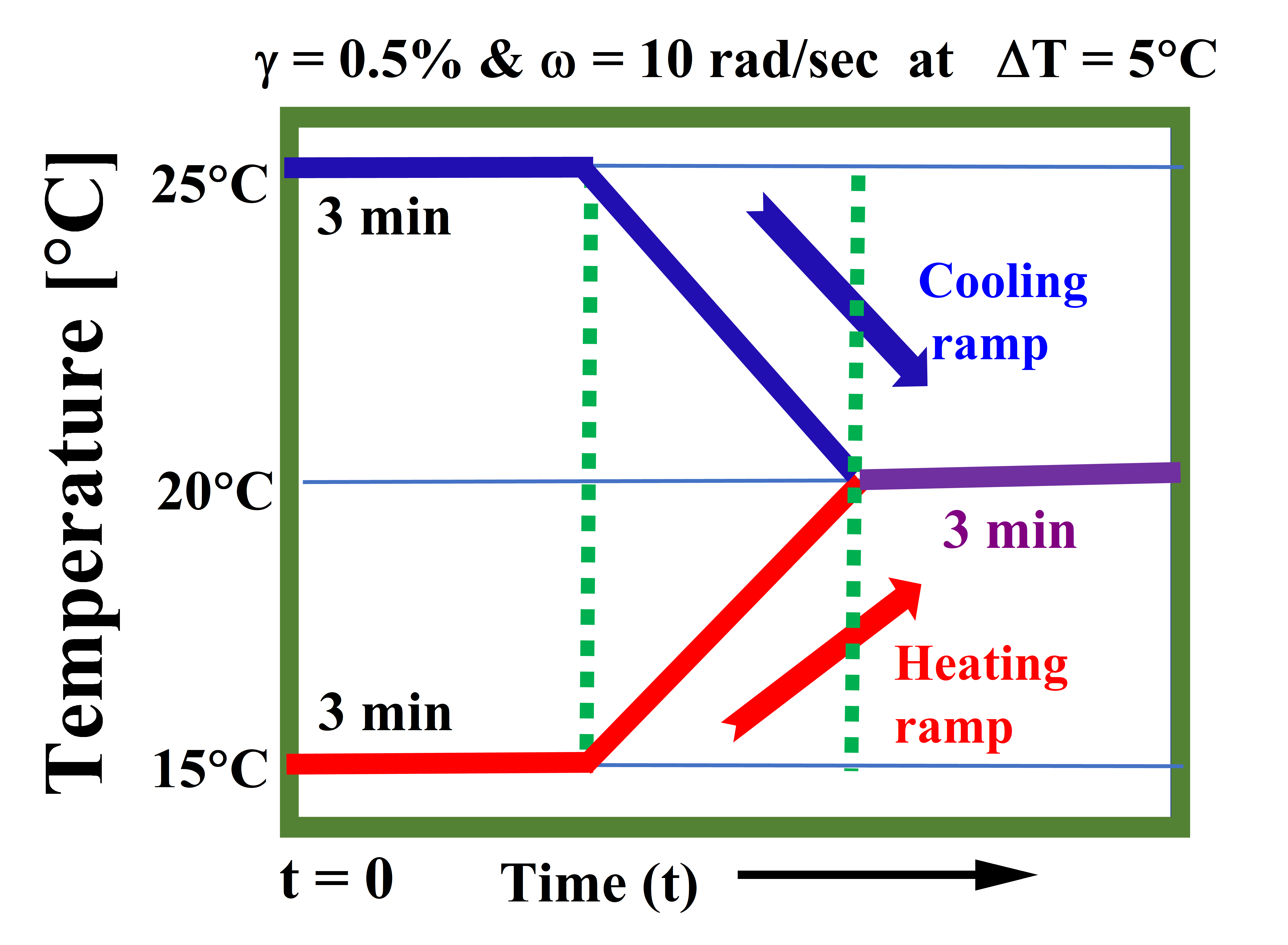}
    \caption{Schematic of the experimental protocol used for heating and cooling temperature ramp experiments. } 
    \label{fig:schematicexp}
\end{figure*}

Figure~\ref{fig:schematicexp} illustrates the experimental protocol used for the temperature ramp Kovacs asymmetry measurements \cite{kawale2026thermal}. All samples were initially heated to 35$^\circ$C to achieve a homogeneous, liquid-like state before applying heating and cooling temperature ramps. Each sample was subsequently cooled to 15$^\circ$C and equilibrated for at least 5 minutes. The time-dependent storage (G$^{\prime}$) and the loss (G$^{\prime\prime}$) moduli were measured at 15$^\circ$C for 3 minutes, after which a temperature ramp was applied to heat the sample to a target temperature of 20$^\circ$C. This procedure is referred to as the heating ramp experiment, and is indicated by the red arrow in Figure~\ref{fig:schematicexp}. The sample at 20$^\circ$C was next heated again to 35$^\circ$C where it liquefied, and then cooled to 25$^\circ$C. The storage and loss moduli were measured at 25$^\circ$C for 3 minutes. Subsequently, a cooling ramp experiment was implemented (indicated by the blue arrow in Figure~\ref{fig:schematicexp}), wherein the sample was cooled to 20$^\circ$C using the same temperature ramp rate as in the heating ramp experiment. The time-dependent moduli were recorded during the heating and cooling ramps. The temperature gradient, $\Delta$T, was therefore maintained at 5$^{\circ}$C during both ramps. After the conclusion of ramp application, the moduli were recorded for an additional 3 minutes. The two green dashed lines in Figure~\ref{fig:schematicexp} indicate the start and end points of the applied temperature ramp. Data for a representative ramp rate of 10$^{\circ}$C/min are presented in the main manuscript, and the data for all other ramp rates are presented in the supporting information.

\subsection{Cryogenic field emission scanning electron microscopy (cryo-FESEM)}
The microstructural analysis of the suspension was performed using a cryogenic field emission scanning electron microscope (cryo-FESEM). After preparing the suspension at a controlled temperature (25$^{\circ}$C), it was placed into a rivet attached to a cryo-stub and rapidly cryo-frozen to -$190$$^{\circ}$C by immersing in liquid nitrogen slush under atmospheric pressure. The vitrified sample was subsequently transferred to the preparation chamber using a cryo-transfer holder. Sublimation was initially carried out for $5$ minutes at -$90^{\circ}$C, followed by fracturing the specimen with a stainless steel knife. To remove residual encapsulated ice, the fractured sample was sublimated for an additional $15$ minutes at -$90^{\circ}$C. A thin platinum coating was applied to the sample under vacuum, using a PP3000T cryo-transfer system (Quorum Technologies). Micrographs were captured using a field-emission scanning electron microscope (FESEM, Ultra Plus FESEM-4098; Carl Zeiss) operated at 5 KeV, with image reconstruction based on back-scattered secondary electron signals.

\begin{figure}
    \centering
    \includegraphics[width=0.35\linewidth]{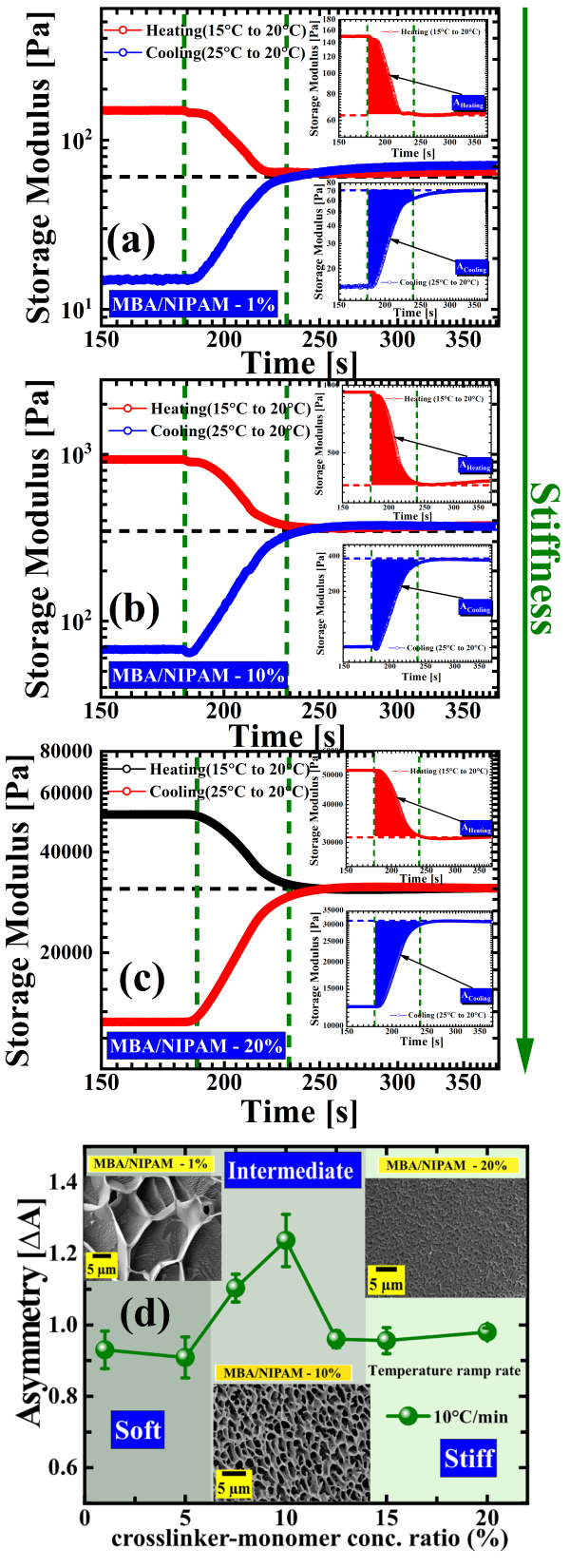}
    \caption{Time-dependent storage moduli (G${^\prime}$) of dense aqueous PNIPAM microgel suspensions ($\phi_{eff}$ = 1.6 prepared at 25$^{\circ}$C), constituted by particles of three representative stiffnesses: (a) 1\%, (b) 10\%, and (c) 20\% {MBA}/{NIPAM}, respectively, during heating and cooling temperature ramps applied at 10$^{\circ}$C/min. The green arrow points toward increasing particle stiffness. Insets in (a)–(c) show the areas under the curves of the storage modulus-time plots during heating (shaded in red) and cooling (shaded in blue) ramps. (d) Corresponding asymmetry of approach ($\Delta$A) toward $T_2$, as a function of microgel stiffness, with the soft, intermediate, and stiff regimes represented by distinct shades of green. Here $\Delta$A = $A_{Heating}$/$A_{Cooling}$, where $A_{Heating}$ and $A_{Cooling}$ are the areas under the storage modulus versus time curves during heating and cooling ramps, respectively. The error bars correspond to standard deviations estimated from three independent measurements. The insets in (d) display cryo-FESEM micrographs of dense suspensions ($\phi_{eff}$ = 1.6, prepared at 25$^{\circ}$C) at three representative particle stiffnesses: 1\%, 10\%, and 20\% MBA/NIPAM (left to right). The scale bar in each cryo-FESEM micrograph represents 5$\mu$m. }
    \label{fig:schematicProtocol}
\end{figure}
\maketitle
\section{Results and Discussion}
\subsection{Evolution of storage (G\texorpdfstring{$^{\prime}$}{'}) and loss (G\texorpdfstring{$^{\prime\prime}$}{''}) moduli under temperature ramps applied at different rates} 
\subsubsection{Dependence of G\texorpdfstring{$^{\prime}$}{'} and asymmetry (\texorpdfstring{$\Delta$A}{DeltaA}) on particle stiffness}
Figures~\ref{fig:schematicProtocol}(a-c) display the raw storage modulus, G$^{\prime}$, data for three representative particle stiffnesses at a temperature ramp rate of 10$^\circ$C/min during both heating and cooling ramps. For all the dense suspensions studied here, the asymmetry in approach of the storage modulus toward the final temperature, $T_2$, during heating and cooling ramps was quantified using an asymmetry parameter, $\Delta$A. This parameter, which characterizes the non-mirror symmetric relaxation of G${^\prime}$ toward $T_2$, is defined as the ratio of the areas under the storage modulus versus time (G${^\prime}$ \textit{vs.} t) curves during the heating and cooling processes~\cite{kawale2026thermal}. For each process, the area under the G${^\prime}$ \textit{vs.} t curve is calculated over the interval corresponding to the applied temperature ramp. The intervals are highlighted in the insets of Figures~\ref{fig:schematicProtocol}(a-c) by two vertical dashed green lines that indicate the start and end of the applied ramps. A horizontal dashed black line, corresponding to the steady state value of G${^\prime}$ at the target temperature $T_{2}$ and extended across the entire ramp, serves as a baseline, as indicated by dashed lines (red for heating, blue for cooling, insets of Figures~\ref{fig:schematicProtocol}(a-c)). The shaded regions denote the areas under the G$^{\prime}$-t curves, with red denoting the heating ramp and blue the cooling ramp. Figure~\ref{fig:schematicProtocol}(d) presents the asymmetry parameter, $\Delta$A, for a representative ramp rate of 10$^\circ$C/min, for suspensions of particles across stiffnesses.  Raw data for the storage moduli across all stiffnesses and ramp rates are provided in Figure~S3 of the supporting information. Variations in the asymmetry parameter $\Delta$A, computed at different ramp rates and plotted versus particle stiffness, are shown in Figure~S4 of the supporting information. The ramp rate determines the volume accessible for particle dynamics during heating and cooling, strongly influencing the magnitude of the resulting asymmetry \cite{kawale2026thermal}. Strikingly, intermediate-stiffness PNIPAM suspensions display maximum structural asymmetry regardless of the applied ramp rate.

The observed asymmetry in G$^{\prime}$ arises because the microgel network traverses fundamentally different energy landscapes along each thermal path. Upon heating, particle deswelling increases the local accessible volume, facilitating faster relaxation of the storage modulus. In contrast, cooling drives particle swelling and tighter packing, thereby restricting the accessible volume and leading to slower, more complex relaxation dynamics. The disparity in structural relaxation between heating and cooling, which dictates the observed evolution of G$^{\prime}$ along both thermal paths, depends strongly on particle stiffness. Soft, loosely crosslinked particles deform readily and undergo substantial swelling and deswelling, preserving sufficient accessible volume across both thermal states. Consequently, the system relaxes toward $T_2$ with similar efficiency along both the heating and cooling paths, thereby reducing path-dependent contrasts (asymmetry) in G$^{\prime}$. Stiff, densely crosslinked particles, on the other hand, experience crowding due to the limited swelling-deswelling capacity of individual particles and remain kinetically jammed throughout the thermal ramp, thereby restricting the accessible volume in both states, also suppressing asymmetry. Furthermore, maintaining a uniform effective volume fraction at 25$^{\circ}$C required us to account for stiffness-dependent particle swelling. Our preparation protocol, therefore, necessitated varying particle numbers across suspensions (Figure~\ref{fig:schematicphieffective}, Table~\ref{table:Synthesis}). This condition reinforced the crowding and jamming of stiff particles, due to the larger particle numbers required to prepare their suspensions. Remarkably, pronounced asymmetry is observed only in intermediate-stiffness particle suspensions. In this regime, particles are neither too soft to fluidize nor too stiff to remain jammed. The disparity in accessible volume between swollen and de-swollen configurations across cooling and heating states produces a maximally asymmetric G$^{\prime}$ response.

PNIPAM suspensions comprising particles of three representative stiffnesses were prepared under quiescent conditions at 25$^\circ$C. Cryo-FESEM images of these samples, acquired after cryo-freezing and presented in the inset of Figure~\ref{fig:schematicProtocol}(d), further validate this picture. Soft particle suspensions generate more open networks (left-most inset panel) due to their lower particle number, thereby increasing accessible free volume. As previously noted, the greater swelling–deswelling capacity of these particles contributes to reduced asymmetry. This observation is consistent with the open network structure presented in Figure~\ref{fig:schematicProtocol}(d). Suspensions comprising intermediate-stiffness particles exhibit more disordered, crowded structures (middle inset panel), where neighboring particles constrain particle dynamics. Upon heating, particle collapse creates volume accessible for motion, whereas subsequent cooling induces particle swelling, restricting this volume. Because heating and cooling induce distinct structural pathways, the suspension responds differently to each thermal ramp, resulting in the pronounced asymmetry observed. Conversely, stiff particle suspensions are highly jammed, characterized by higher particle number densities (right-most inset panel). In these systems, particles remain caged by their neighbors during both heating and cooling, leading to minimal structural evolution and suppressed asymmetry. We conclude that the observed Kovacs asymmetry serves as an indicator of structural memory, with the path-dependent relaxation of the storage modulus governed by both the instantaneous structural state and thermal history of the samples.

\subsubsection{Estimation of energy dissipation across applied ramp rates}
\begin{figure}[!t]
    \centering
    \includegraphics[width=0.95\linewidth]{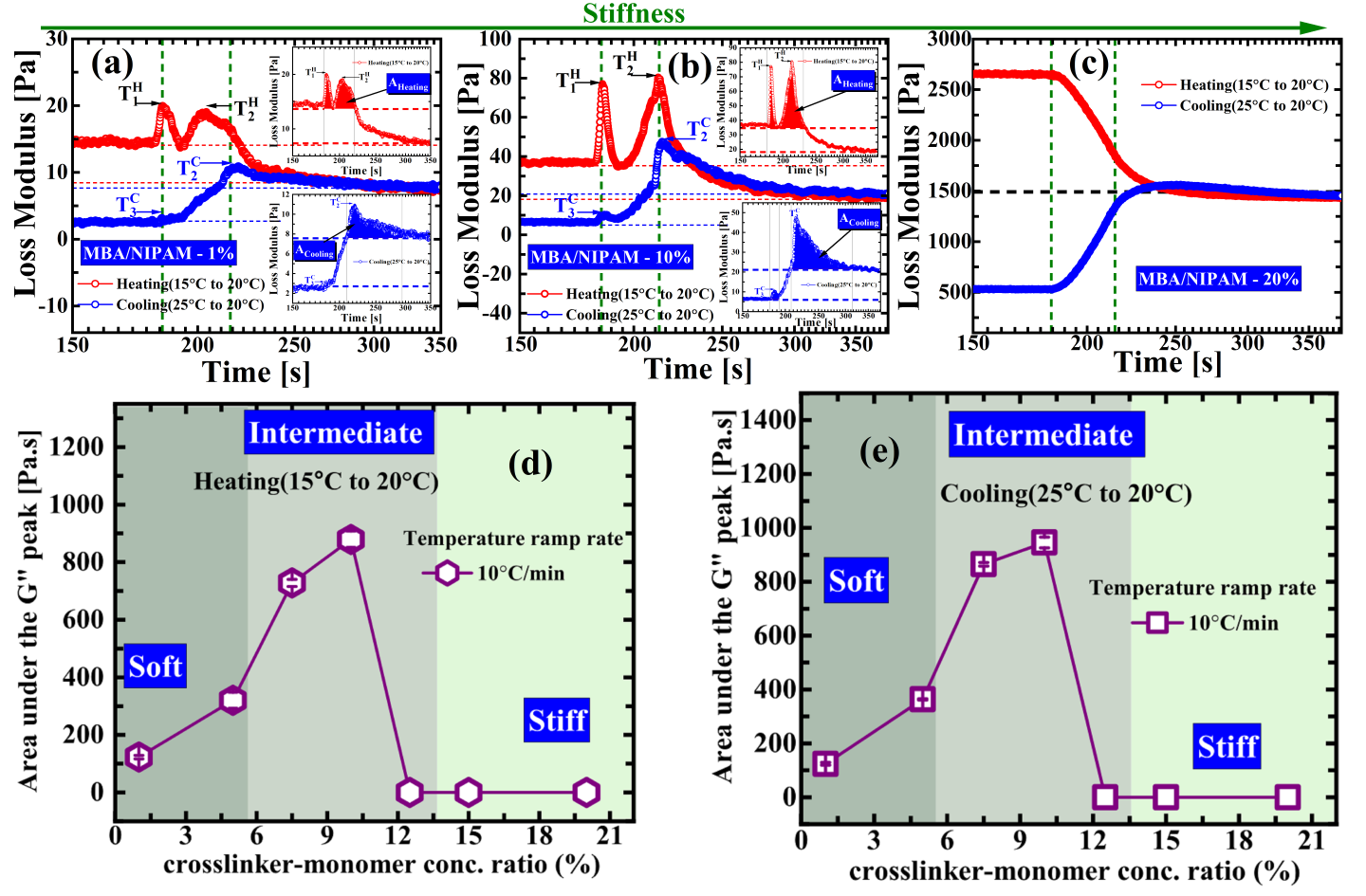}
    \caption{Time-dependent loss moduli (G$^{\prime\prime}$) of dense aqueous PNIPAM microgel suspensions ($\phi_{eff}$ = 1.6 prepared at 25$^{\circ}$C) during heating and cooling temperature ramps at rate of 10$^{\circ}$C/min, for three representative particle stiffnesses: (a) 1\%, (b) 10\%, and (c) 20\% MBA/NIPAM, respectively. Insets in (a) and (b) show the area under the curves of the loss modulus-time plots during the heating (shaded in red) and cooling (shaded in blue) ramps. The green arrow points in the direction of the increasing particle stiffness. Areas under the loss modulus (G$^{\prime\prime}$-t) peaks during (d) heating and (e) cooling ramps at a rate of 10$^{\circ}$C/min as a function of microgel stiffness,  with the soft, intermediate, and stiff regimes represented by distinct shades of green. The error bars were calculated from three independent measurements.}
    \label{fig:schematicLMallrates}
\end{figure}

Figures~\ref{fig:schematicLMallrates}(a-c) present the loss modulus response of dense suspensions comprising PNIPAM particles of three representative stiffnesses: 1\%, 10\%, and 20\% MBA/NIPAM, acquired during heating and cooling ramps applied at 10$^\circ$C/min. The loss modulus, G$^{\prime\prime}$, response is characterized by dissipation peaks~\cite{mason1996yielding,menut2012does} at the start and end of each temperature ramp. These peaks are clearly seen for the soft and intermediate-stiffness particle suspensions (Figures~\ref{fig:schematicLMallrates}(a-b)) at the higher ramp rates, but are absent for the stiffer particle suspensions (Figure~\ref{fig:schematicLMallrates}(c)). G$^{\prime\prime}$ data corresponding to other particle stiffnesses, acquired during both heating and cooling ramps, show the same trend: peaks are present for soft and intermediate particle suspensions in all experiments except at the lowest ramp rate, but disappear for stiff particle suspensions (Figures~S5 and ~S6 of the supporting information).

In our earlier work~\cite{kawale2026thermal}, we had shown that G$^{\prime\prime}$ peaks arise from microgel rearrangements triggered by thermal shocks at the start and end of a thermal ramp, with the peak height reflecting the magnitude of energy dissipation and hence the intensity of rearrangements. Here, we examine how energy dissipation due to rearrangement events depends on particle stiffness. We quantify the extent of energy dissipation by monitoring the area under the G$^{\prime\prime}$-t peak, which measures the cumulative energy dissipated during the ramp and thus the overall extent of structural rearrangement. These areas were computed relative to a horizontal baseline, as indicated by the dashed lines in the insets of Figures~\ref{fig:schematicLMallrates}(a) and~\ref{fig:schematicLMallrates}(b) (red for heating, blue for cooling). For heating ramps, the baseline corresponded to the G$^{\prime\prime}$ plateau value before the ramp onset, whereas for cooling ramps, it was defined as the G$^{\prime\prime}$ plateau value after the ramp conclusion. In both scenarios, the baseline was extended horizontally across the entire temperature ramp. The area enclosed between the peak and its baseline, shaded red for heating and blue for cooling in the insets, quantifies the energy dissipated during the corresponding rearrangement event. The analysis was carried out independently for the peaks observed during the heating and cooling ramps. The areas thus computed are plotted against particle stiffness in Figures~\ref{fig:schematicLMallrates}(d) and~\ref{fig:schematicLMallrates}(e) for the heating and cooling ramps, respectively. Interestingly, the calculated areas, and therefore the cumulative dissipation, vary non-monotonically with particle stiffness, reaching their maximum for suspensions of intermediate-stiffness particles. 
This non-monotonic dependence on particle stiffness can be rationalized by considering how the volume available for particle dynamics changes during heating (deswelling) and cooling (swelling), which in turn affects the resistance that particles experience from their neighbors. Soft particles, owing to their high compliance, absorb the imposed thermal changes primarily through local deformation, ensuring minimal translation relative to their neighbors. This mechanism results in modest dissipation. Stiff particles undergo minimal swelling-deswelling, which keeps the suspension in a kinetically arrested, densely packed state throughout the ramp. The restricted accessible volume limits structural reorganization, so dissipation remains low. Particles of intermediate stiffness swell and shrink considerably, but their limited deformability prevents them from accommodating the resulting local stresses. They therefore rearrange by moving past one another, leading to the maximum energy dissipation and the largest G$^{\prime\prime}$ peak area.

Our experiments across different ramp rates (Figures~S5-S7) revealed the same non-monotonic dependence of energy dissipation on stiffness, although the variations were ramp rate-dependent. This behavior is expected since the ramp rate dictates the kinetics of particle rearrangements, consistent with our previous findings~\cite{kawale2026thermal}. Notably, at the slowest ramp rate of 1$^\circ$C/min, the G$^{\prime\prime}$ response showed no distinguishable peaks (Figures~S5(a) and ~S6(a)) as the particles had sufficient time to relax. Cryo-FESEM micrographs (insets of Figure~\ref{fig:schematicProtocol}(d)) support these findings, revealing morphologies that are consistent with the extents of rearrangement inferred from the G$^{\prime\prime}$ peak areas. In the soft suspension, particles can deform easily but do not rearrange, so the network retains its open, connected structure. In the stiff particle suspension, the jammed packing leaves little room for particles to move, so the arrested configuration is preserved throughout the ramp. In contrast, the disordered, compact structure of the intermediate-stiffness particle suspension allows particles to move past one another and settle into new positions during the ramp. Together with the G$^{\prime\prime}$ peak areas, these morphological differences confirm that the rearrangements and extent of structural disorder are both maximum in intermediate-stiffness particle suspensions. 

Thermally induced dissipation in dense PNIPAM suspensions is therefore non-monotonic in particle stiffness and is governed by the balance between swelling capacity and accessible volume within the packing. This non-monotonic behavior is not limited to the G$^{\prime\prime}$ peak area. As expected, the G$^{\prime\prime}$ peak heights (Section~3 of the supporting information) follow the same trend, with a maximum at intermediate stiffness (Figure~S8). Therefore, the G$^{\prime\prime}$ peak area, which quantifies the energy dissipated during particle rearrangements, and the G$^{\prime}$ asymmetry, $\Delta$A, which reflects the thermal-path dependence of the resulting structure, are both maximal at intermediate stiffness. This coincidence indicates that the rearrangements that dissipate energy are also those that render the structure path-dependent. Such a correlation between dissipation and history-dependent structural recovery is reminiscent of glassy, out-of-equilibrium systems and suggests that microgels of intermediate stiffness are the most promising candidates for forming glasses with tunable, history-dependent properties.

 \subsection{Maximal asymmetry and dissipation in intermediate-stiffness particle suspensions when particle number density is fixed across systems}
 Because particle stiffness governs swelling and deswelling capacity, preparing suspensions of equal effective volume fractions necessitates different particle concentrations for each particle stiffness. To decouple stiffness from the suspension preparation protocol, we repeated the measurements at equal particle number density. The Kovacs asymmetry and rearrangement-driven energy dissipation remained maximal at intermediate stiffness across all ramp rates (raw data in Figures~S9 and~S10; analysis in Figures~S11 and~S12 of the supporting information), indicating that the observed non-monotonic trend is intrinsic to particle stiffness rather than the suspension preparation protocol. The contrast between intermediate and extreme stiffnesses is weaker at equal number density because the effective volume fraction is no longer matched. 
 
 Stiff particles are less densely packed in equal number density systems when compared to the earlier samples prepared at $\phi_{eff}$ = 1.6 at 25$^{\circ}$C, and retain more accessible volume for rearrangement. The soft particle suspensions prepared \textit{via} this protocol are more crowded, thereby narrowing the differences among the three stiffnesses (Figure~S13 of the supporting information). In soft particle suspensions, hydrophobic coagulation generates a gel-like network that persists well below the VPTT \cite{misra2024effect}. Heating drives particle collapse, whereas cooling-induced reswelling is constrained by interparticle attraction and particle crowding. Consequently, configurational changes are less significant during cooling and heating, thereby reducing asymmetry. The inherent softness of these particles also enables them to adapt to neighboring particles with minimal rearrangement, which reduces energy dissipation. In stiff particle suspensions, the limited capacity for swelling and deswelling restricts configurational changes, thereby suppressing both asymmetry and energy dissipation despite the presence of some accessible volume for rearrangement. Memory effects are therefore consistently most pronounced at intermediate stiffness, where particles can swell and rearrange without undergoing coagulation or jamming.

\maketitle
\section{Conclusions}
 Our findings show that path-dependent structural recovery in dense PNIPAM suspensions is governed by particle stiffness, which was tuned by varying the crosslinker concentration during synthesis. Oscillatory rheology experiments were performed to study the relaxation of the viscoelastic moduli under heating and cooling temperature ramps applied at varying rates. The Kovacs asymmetry, obtained from the storage modulus (G$^{\prime}$) response, and the associated dissipation, obtained from the loss modulus (G$^{\prime\prime}$) response, were both maximal for suspensions of intermediate-stiffness particles. This non-monotonic dependence on particle stiffness arises from the stiffness-dependent swelling and deswelling kinetics of the particles and the volumes available for their rearrangement. Soft microgels deform, swell, and deswell substantially, accommodating imposed temperature ramps by changing their size and shape. This limits rearrangements, leading to minimal path dependence and reduced asymmetry. Stiffer particles swell only weakly and deform little, which keeps the suspension densely packed and limits structural rearrangements, also suppressing asymmetry. Particles of intermediate stiffness swell moderately while retaining partial resistance to compression. Deswelling upon heating and re-swelling upon cooling therefore drive the system through distinct regions of the energy landscape, producing non-mirror-symmetric structural recovery and the largest Kovacs asymmetry, along with maximum dissipation. Particle stiffness thus emerges as a tunable parameter for controlling memory and path-dependent relaxation in soft glassy materials.

Cryo-FESEM imaging reveals that particle stiffness governs both network architecture \cite{misra2024effect} and thermal response. Soft particles form open, dilute networks that accommodate temperature fluctuations through efficient swelling and deswelling. In contrast, stiff particles form crowded, rigid networks that restrict particle-scale dynamics and suppress thermal reorganization. As a result, Kovacs asymmetry and dissipation are minimal at both extremes of stiffness. Particles with intermediate stiffness form disordered networks that exhibit moderate swelling and undergo dynamic rearrangement, thereby maximizing both structural and dynamic responses. Therefore, stiffness governs particle self-assembly, which in turn regulates energy dissipation and structural recovery. These trends persist irrespective of whether the system is prepared at a constant effective volume fraction or a constant particle number density.

PNIPAM particles of intermediate stiffness, therefore, clearly provide an optimal balance between swelling capability and accessible volume in dense packings, resulting in the most extensive particle rearrangements. 
Structural recovery and memory-like behavior can therefore be controlled through particle stiffness. This makes intermediate-stiffness PNIPAM microgels particularly promising candidates as colloidal glasses, while particle stiffness emerges as a useful parameter for studying structural relaxation and non-equilibrium dynamics in microgel suspensions and other soft glassy systems. PNIPAM microgels, widely used in drug delivery, tunable rheological fluids, and responsive coatings, often undergo repeated thermal cycling during processing and application \cite{franco2025soft, guan20243d, throat2024macromolecular, nigro2024influence}. Controlling stiffness may allow us to design materials with tunable history dependence or deliberately-enhanced path-encoded memory.

\section*{Author contributions}
\textbf{Sonali Vasant Kawale:} Data Curation, Methodology, Validation, Formal analysis, Investigation, Writing - Original Draft, Visualization. 
\textbf{Ranjini Bandyopadhyay:} Conceptualization, Methodology, Validation, Resources, Data curation, Visualization, Writing - Review \& Editing, Supervision, Project administration, Funding acquisition.

\section*{ACKNOWLEDGMENTS}
The authors thank K. M. Yatheendran for his help with cryo-FESEM imaging and acknowledge Yogesh M. Joshi for useful discussions during the preliminary stage of this work. 

\section*{Conflicts of interest} 
The authors have no conflict of interest to disclose.

\renewcommand\refname{References}
	
	\bibliographystyle{elsarticle-num}
	\bibliography{Reference}
    

\section*{Supplementary material}

# 1 Average hydrodynamic diameters and temperature-dependent swelling ratios of PNIPAM particles of different stiffnesses in aqueous suspension

Dynamic light scattering (DLS) measurements were performed using a Brookhaven Instruments Corporation (BIC) BI-200SM spectrometer. Detailed descriptions of the experimental setup are provided elsewhere [1] [2]. Temperature regulation of the PNIPAM suspensions, ranging from 12°C to 55°C, was maintained with a Polyscience digital water-circulating temperature controller. The intensity autocorrelation function, $g^{(2)}(q,t)$, of the scattered light was measured using a Brookhaven Instruments digital autocorrelator (BI-9000AT). The function, $g^{(2)}(q,t)=\frac{\langle I(q,0)I(q,t)\rangle}{\langle I(q,0)\rangle^2}= 1+\mathrm{A}|g^{(1)}(q,t)|^2$ [3], where I($q$,$t$) denotes the scattered intensity at scattering wave vector $q$ and delay time $t$, $g^{(1)}(q,t)$ is the normalized electric field autocorrelation function, and $A$ is the coherence factor. The angular brackets $\langle\rangle$ signify temporal averaging. The scattering wave vector, $q = (4\pi \mathrm{n}/\lambda)\ \sin(\theta/2)$, where scattering angle $\theta = 90°$, refractive index of the medium $n \approx 1.33$, and laser wavelength $\lambda = 532$ nm. For suspensions of PNIPAM particles with size polydispersity, the decay of the normalized autocorrelation function, $C(t) = \frac{g^{(2)}(q,t)-1}{A}$ [4], was modeled using a stretched exponential: $C(t) = [e^{-(t/\tau)^{\beta}}]^2$. Figure S1(a) shows $C(t)$ as a function of $t$ at different temperatures for dilute suspensions of PNIPAM particles prepared with a 10% crosslinker-monomer concentration ratio and fits to a stretched exponential function. The mean relaxation time, $\langle\tau\rangle = (\tau/\beta)\Gamma(1/\beta)$, with $\beta$ as the stretching exponent, $\tau$ is the relaxation time, and $\Gamma$ is the Euler Gamma function [5]. The average hydrodynamic diameter, $\langle d_H\rangle$, was extracted using the Stokes-Einstein relation: $\langle d_H\rangle = \frac{k_B T\langle\tau\rangle q^2}{3\pi\eta}$, where $k_B$ is the Boltzmann constant, $T$ the absolute temperature, and $\eta$ the solvent viscosity [4] [5]. Figure S1(b) presents the temperature-dependent swelling ratios of PNIPAM particles of different stiffnesses. To ensure thermal equilibration prior to measurements of hydrodynamic diameters, the samples were equilibrated for 20 minutes at each temperature.

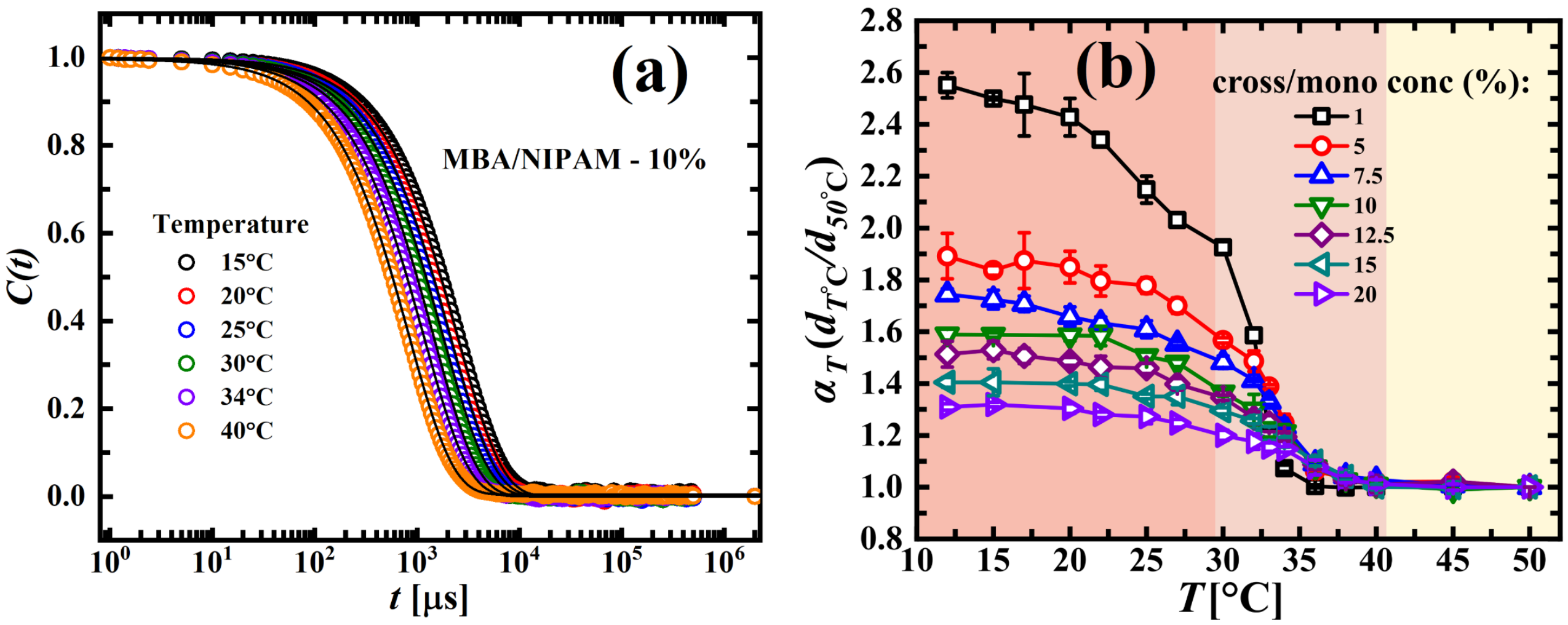


FIGURE S1: (a) The normalized intensity autocorrelation functions, $C(t)$, as a function of delay time, $t$, for dilute suspensions of PNIPAM particles prepared with 10% crosslinker-monomer concentration ratio. The black solid lines are fits to the expression: $C(t) = [\exp(-(\frac{t}{\tau})^{\beta})]^2$. (b) Temperature-dependent swelling ratios, $\alpha_T = (d_{T^\circ\mathrm{C}}/d_{50^\circ\mathrm{C}})$, of PNIPAM particles of different stiffnesses (labeled as cross/mono conc in figure legend) in aqueous suspension. Here, $d_{T^\circ\mathrm{C}}$ and $d_{50^\circ\mathrm{C}}$ are the average hydrodynamic diameters of PNIPAM particles at temperatures $T$ and 50°C, respectively.

## 2 Estimation of effective volume fractions of dense suspensions of PNIPAM particles of different stiffnesses

The viscosities ($\eta$) of aqueous PNIPAM suspensions of different particle stiffnesses were measured using rotational rheometry across a broad range of shear rates ($\dot{\gamma} = 0.01$ to $1000\ s^{-1}$). The acquired viscosity-shear rate data were fitted to the Cross model [6]:

$$\frac{(\eta - \eta_\infty)}{(\eta_0 - \eta_\infty)} = \frac{1}{1 + (k\dot{\gamma})^m} \tag{1}$$

In this model, $\eta_0$ and $\eta_\infty$ denote the viscosities at very low (zero-shear) and very high shear rates, respectively. The parameter $k$ represents a time constant associated with microgel relaxation, and $m$ is a dimensionless exponent. These parameters were determined by fitting the experimental $\eta$ versus $\dot{\gamma}$ curves to the Cross model, as illustrated in the inset of Figure S2, which displays representative data for PNIPAM particles with 15% crosslinker-monomer concentration ratio. The relative viscosity ($\eta_{rel}$) was calculated as the ratio of the zero-shear viscosity ($\eta_0$) of each PNIPAM suspension and the viscosity of water ($\eta_s$), $\eta_{rel} = \eta_0/\eta_s$.

For soft, deformable, and compressible particles, the conventional volume fraction ($\phi$) does not

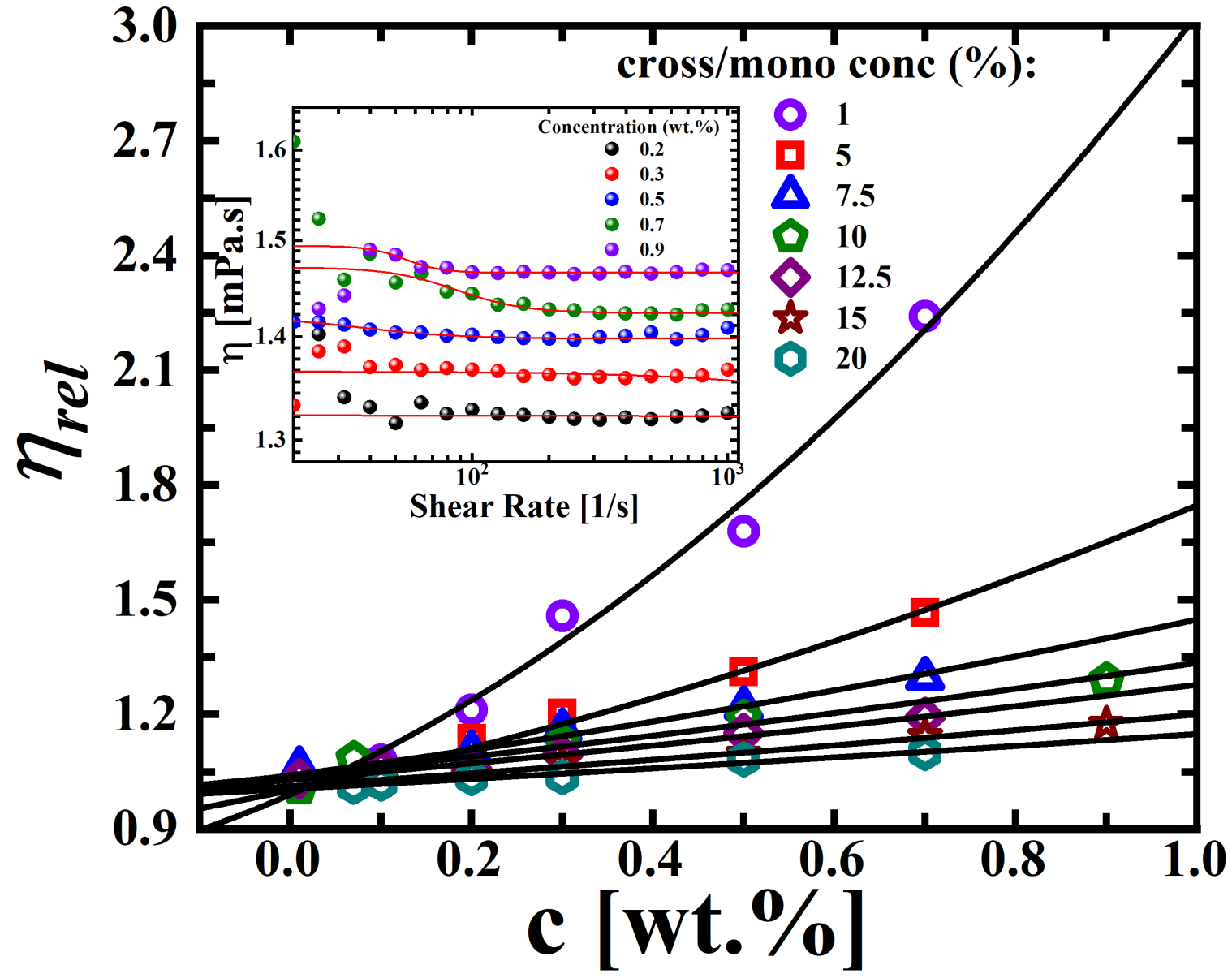


FIGURE S2: Relative viscosity, $\eta_{rel}$, *vs.* particle concentration, c in wt.%, of aqueous suspensions of PNIPAM particles of different particle stiffnesses. The black solid lines are fits to Eqn. 2 of the supporting information. The inset displays representative data for $\eta$ as a function of $\dot{\gamma}$ for suspensions of PNIPAM particles prepared at different concentrations with 15% MBA/NIPAM, and the corresponding fits to the Cross model (Eqn. 1 of the supporting information).

accurately represent particle packing. Therefore, an effective volume fraction was estimated for each suspension. The effective volume fraction is defined as $\phi_{eff} = nV_d$ [7–9], where $n$ is the particle number density per unit volume and $V_d = \frac{\pi(\langle d_H \rangle)^3}{6}$ represents the volume of an undeformed particle of average hydrodynamic diameter $\langle d_H \rangle$ in a dilute suspension. The relative viscosities, $\eta_{rel}$, measured at 25°C, are plotted as a function of particle concentration, $c$, for PNIPAM particles of different stiffnesses and fitted to Batchelor's equation (Figure S2) [10, 11]:

$$\eta_{rel} = 1 + 2.5(c/c_p) + 5.9(c/c_p)^2 \tag{2}$$

The polymer concentration within each particle ($c_p$) was determined by fitting the Batchelor's equation to the data (black lines, Figure S2) acquired for each suspension. Effective volume fractions were calculated using $\phi_{eff} = c/c_p$. The estimated values of $c_p$, $\phi_{eff}$, and the concentrations, $c$, of the PNIPAM particles of different stiffnesses are provided in Table 1 of the main paper.

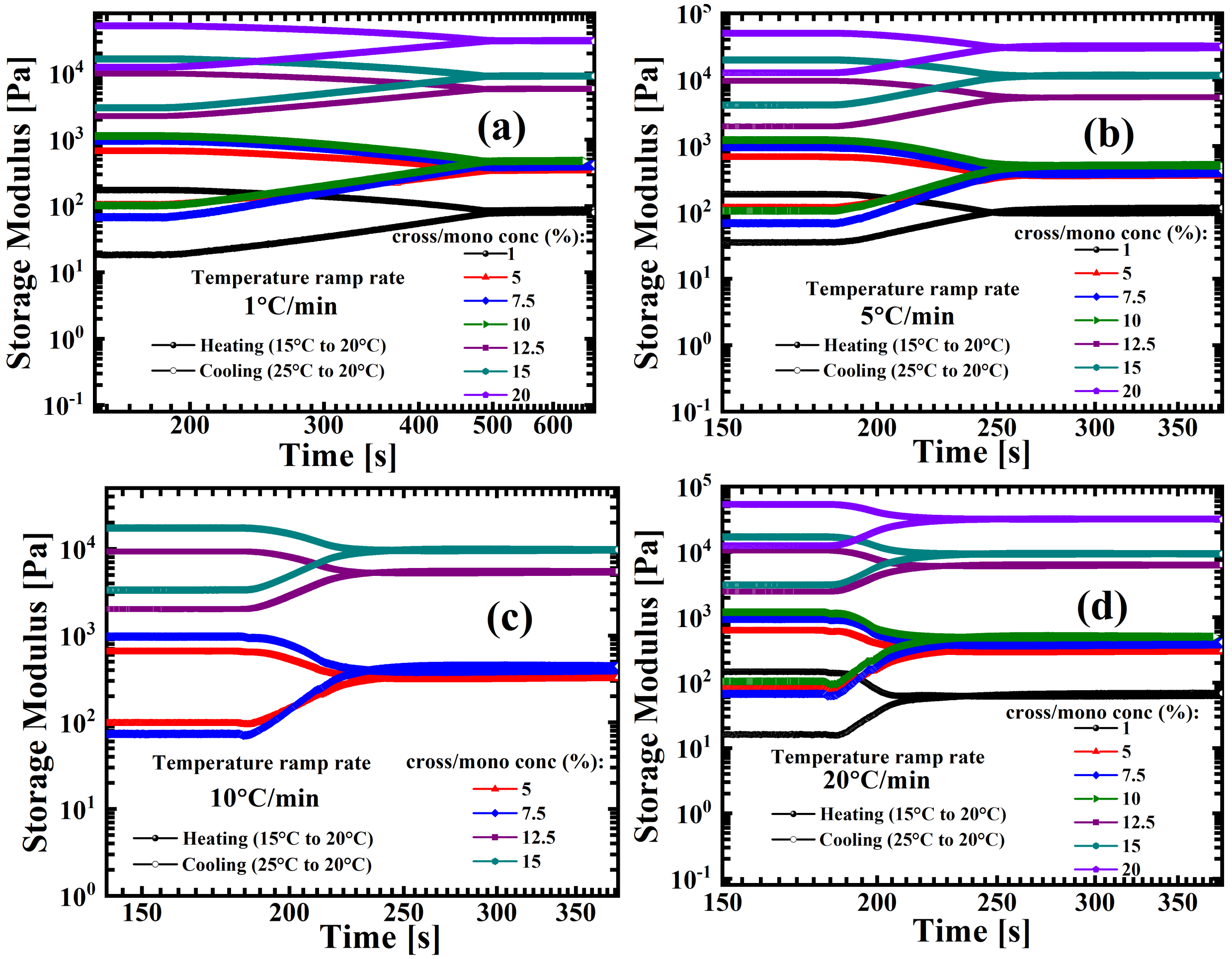


FIGURE S3: Storage modulus, G′, as a function of time during heating and cooling temperature ramps at rates: (a) 1°C/min (b) 5°C/min (c) 10°C/min and (d) 20°C/min for dense aqueous PNIPAM suspensions ($\phi_{eff}$ = 1.6 prepared at 25°C) constituted by particles of different stiffnesses.

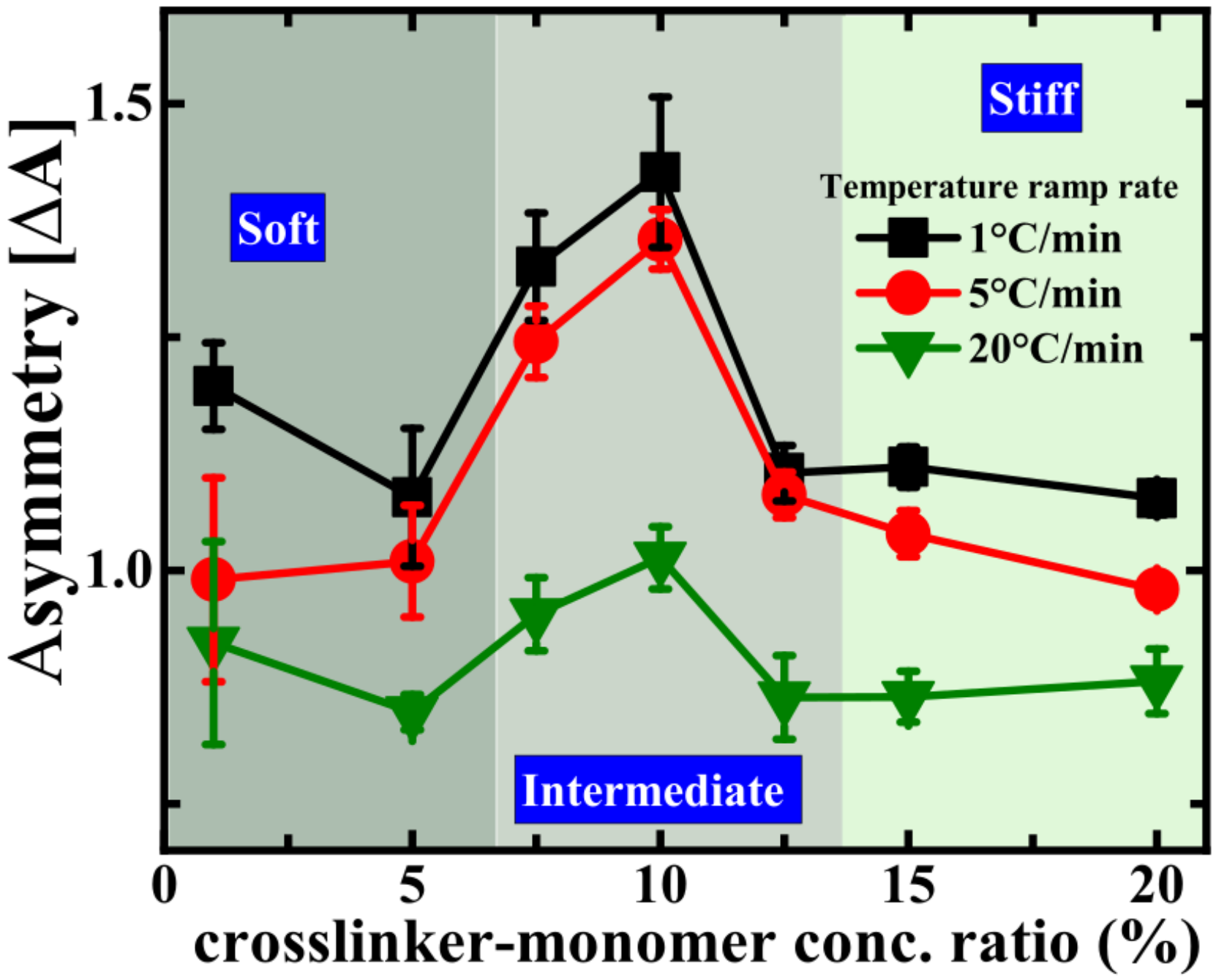


FIGURE S4: Asymmetry of approach (ΔA) towards $T_2$, for dense aqueous PNIPAM suspensions ($\phi_{eff}$ = 1.6 prepared at 25°C) as a function of microgel particle stiffness (crosslinker-monomer conc. ratio (%)), under varying temperature ramp rates.

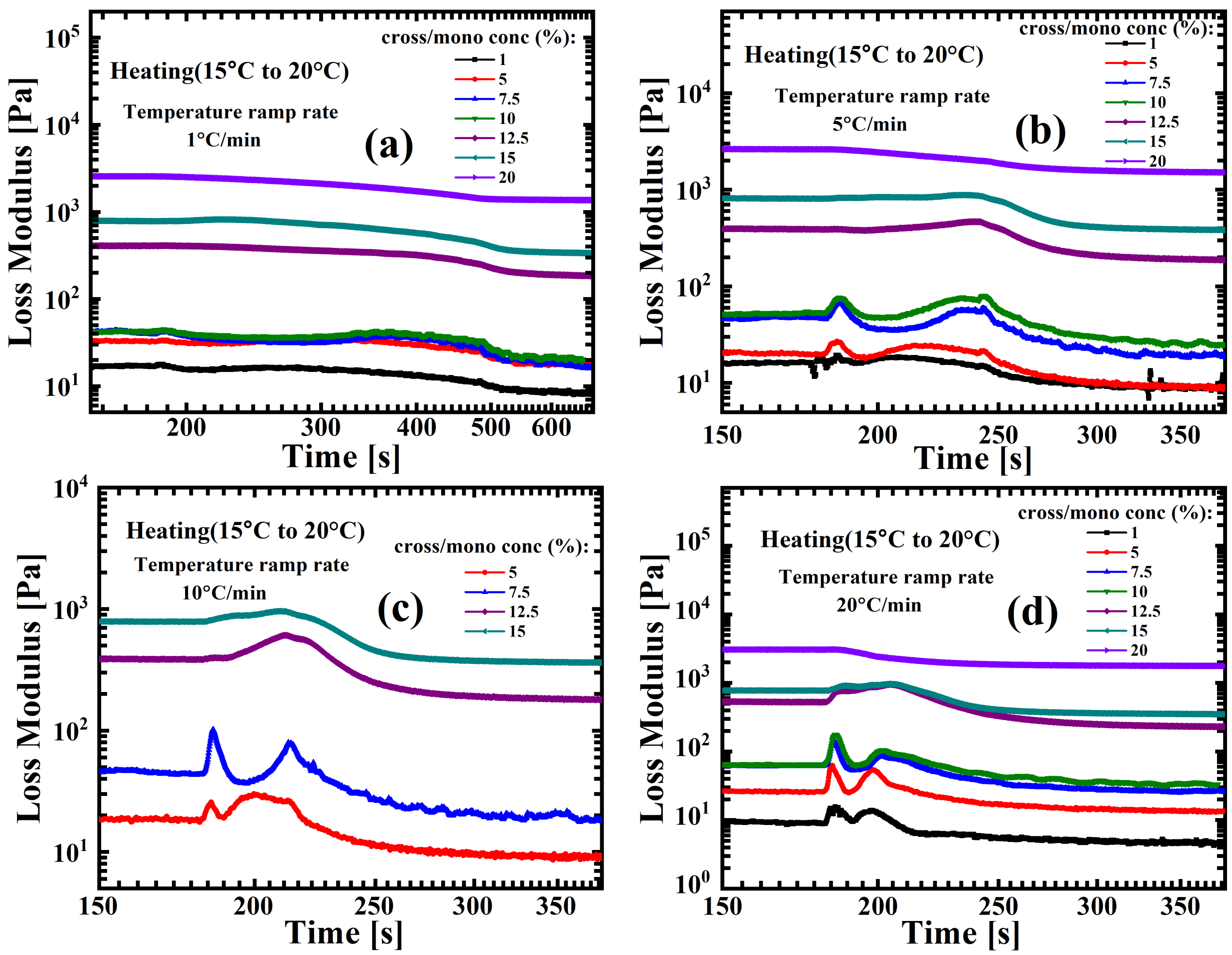


FIGURE S5: Loss modulus, G″, acquired *vs.* time during heating ramps at rates: (a) 1°C/min (b) 5°C/min (c) 10°C/min and (d) 20°C/min, applied to dense aqueous PNIPAM suspensions ($\phi_{eff} = 1.6$ prepared at 25°C) constituted by particles of different stiffnesses.

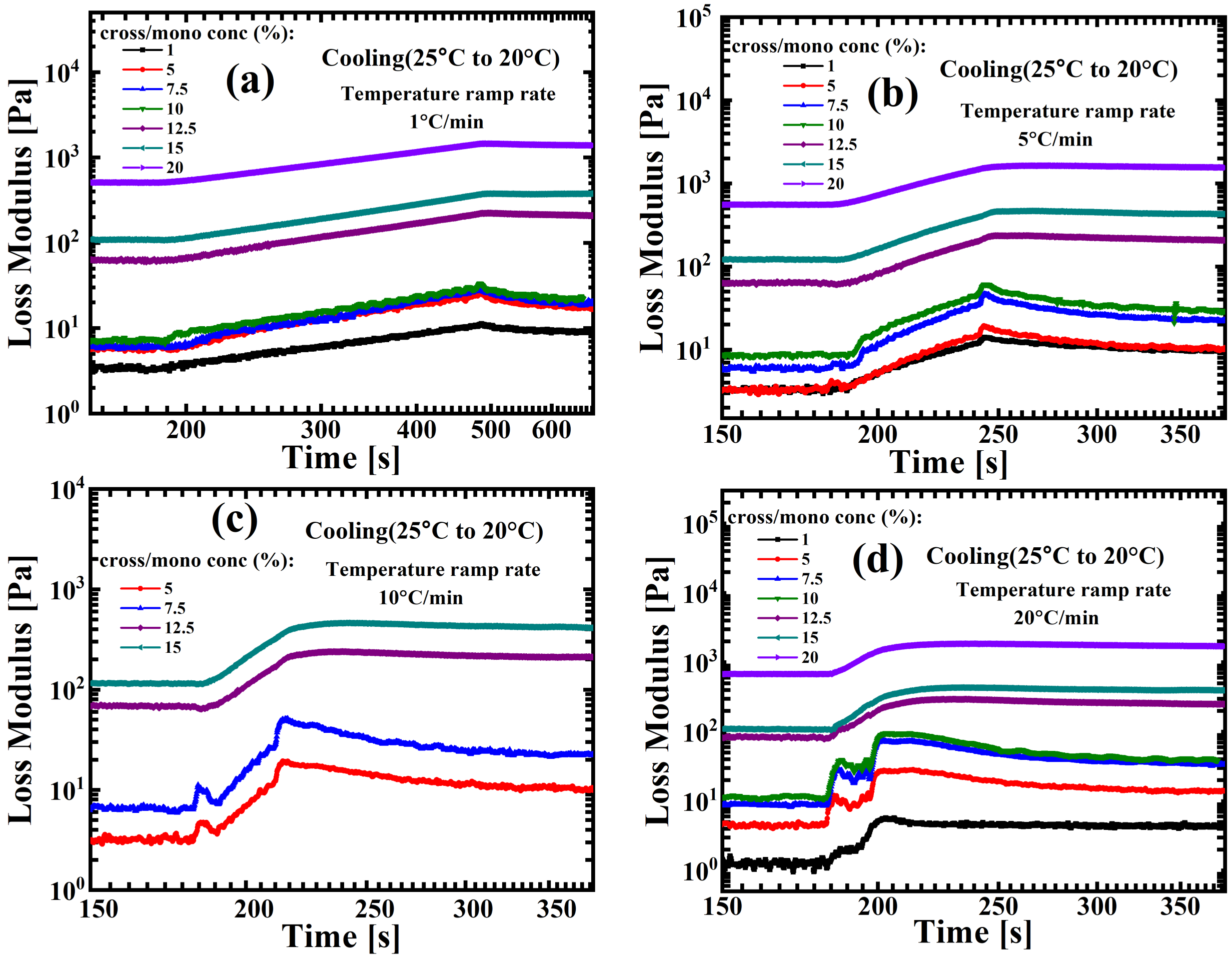


FIGURE S6: Loss modulus, G″, *vs.* time during cooling ramps at rates: (a) 1°C/min (b) 5°C/min (c) 10°C/min and (d) 20°C/min, applied to dense aqueous PNIPAM suspensions ($\phi_{eff}$ = 1.6 prepared at 25°C) constituted by particles of different stiffnesses.

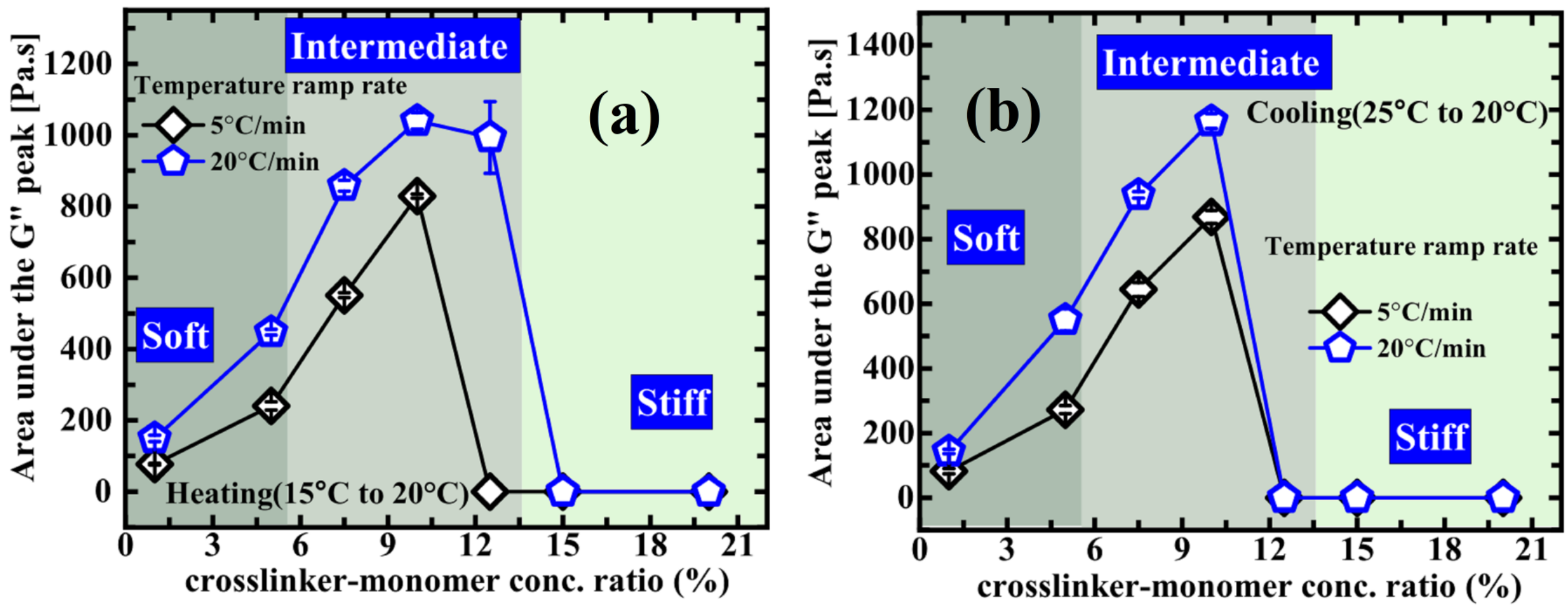


FIGURE S7: The areas under the loss modulus (G″-t) peaks for dense aqueous PNIPAM suspensions ($\phi_{eff}$ = 1.6, prepared at 25°C) during (a) heating and (b) cooling ramps applied at different rates. The data is plotted as a function of microgel stiffness, with the soft, intermediate, and stiff regimes represented by distinct shades of green. The error bars were estimated from three independent measurements.

# 3 Estimation of loss modulus peak heights in suspensions of PNIPAM particles of different stiffnesses

The relaxation of the loss modulus, G″, is characterized by dissipation peaks at the initiation and termination of each temperature ramp, as illustrated in Figures S5 and S6. A horizontal baseline was established as per the protocol described in Section 3.1.2 of the main paper. For heating ramps, the baseline was set to the G″ plateau value prior to ramp onset, while for cooling ramps, it was set to the G″ plateau value after ramp completion. In both cases, the baseline was extended horizontally throughout the entire temperature ramp. The peak heights were then determined by measuring the vertical distance between the G″ value at the peak maximum and the baseline value at the corresponding time point, and plotted in Figure S8.

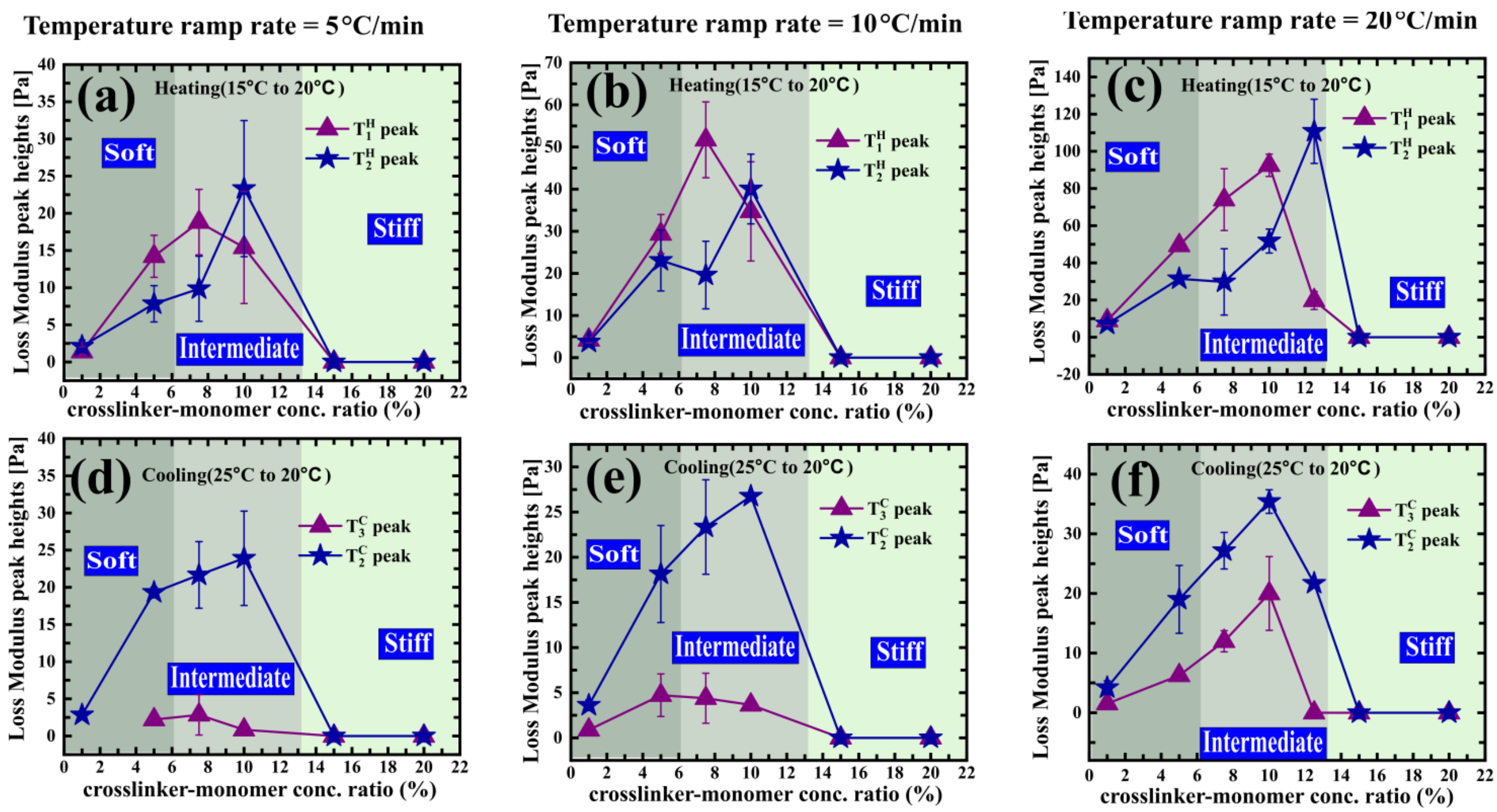


FIGURE S8: Loss modulus, G″, peak heights when dense aqueous PNIPAM suspensions ($\phi_{eff} = 1.6$ prepared at 25°C) of different particle stiffnesses were heated from 15°C to 20°C at temperature ramp rates (a) 5°C/min, (b) 10°C/min, and (c) 20°C/min respectively. G″ peak heights when the same suspensions were cooled from 25°C to 20°C at temperature ramp rates (d) 5°C/min, (e) 10°C/min, and (f) 20°C/min respectively. $T_1^H$ and $T_2^H$ in (a)-(c) represent the loss modulus peaks at the start and end of the heating ramp when the suspension was heated toward $T_2$, while $T_3^C$ and $T_2^C$ in (d)-(f) represent the loss modulus peaks as the suspension was cooled toward $T_2$. The soft, intermediate, and stiff regimes are represented by distinct shades of green. The error bars denote standard deviations estimated from three independent measurements.

# 4 Asymmetry of approach, ΔA, and energy dissipation in suspensions prepared at constant particle number density

To isolate the effect of particle stiffness from preparation-induced variations, experiments were repeated with suspensions of equal particle number density at ramp rates of 1, 5, 10, and 20°C/min. The sample preparation protocol is described in Section 2.3 of the main paper. The raw storage (G′) and loss (G″) moduli data are shown in Figures S9 and S10, and the resulting asymmetry in the approach of the storage modulus toward $T_2$ and the area under the loss modulus peaks are shown in Figure S11. The loss modulus peak heights across all ramp rates are given in Figure S12. Cryo-FESEM micrographs of the corresponding suspension microstructures are provided in Figure S13.

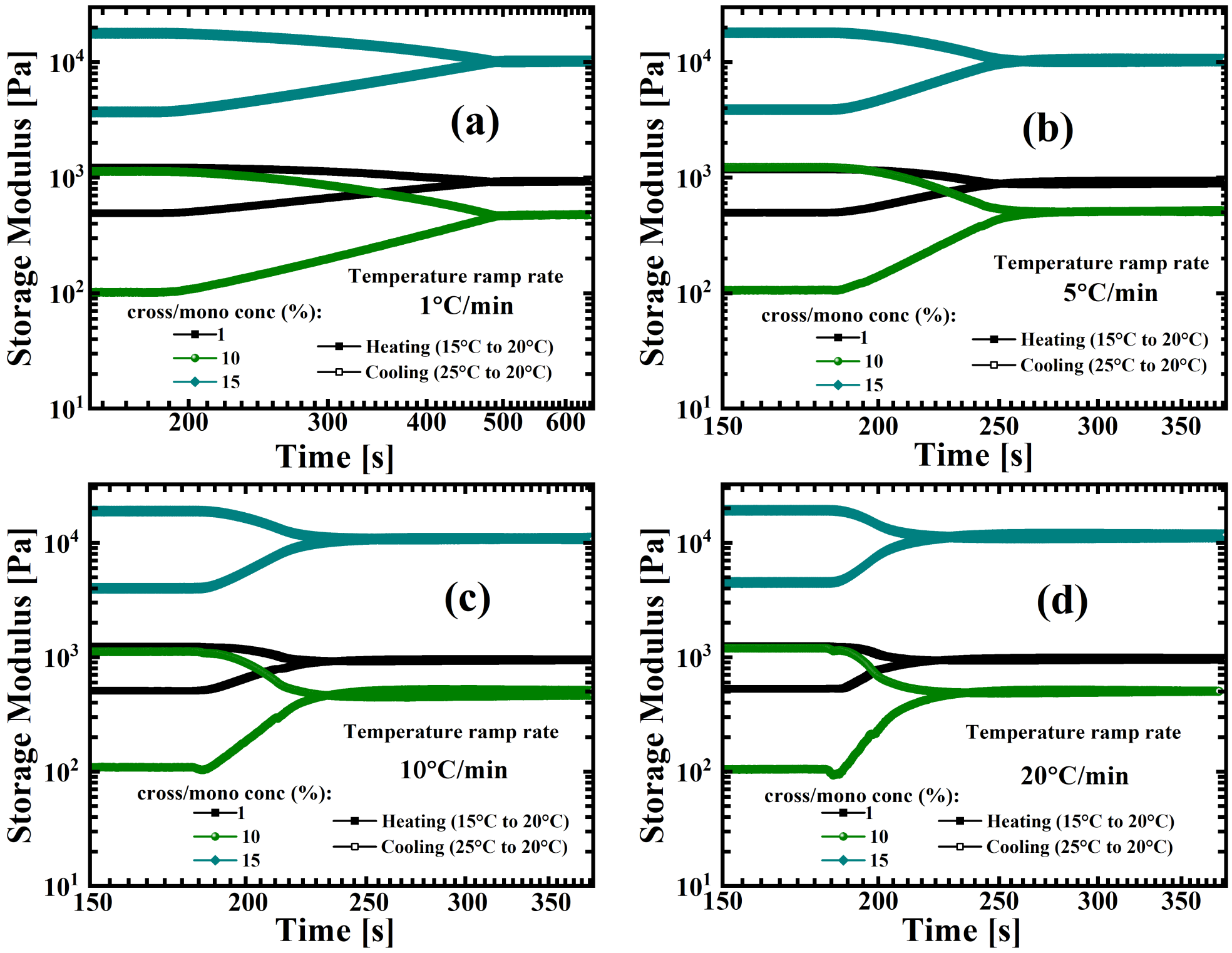


FIGURE S9: Storage modulus, G′, as a function of time during heating and cooling temperature ramps at rates: (a) 1°C/min (b) 5°C/min (c) 10°C/min and (d) 20°C/min, measured for dense PNIPAM suspensions prepared with particles of different stiffnesses at constant number density.

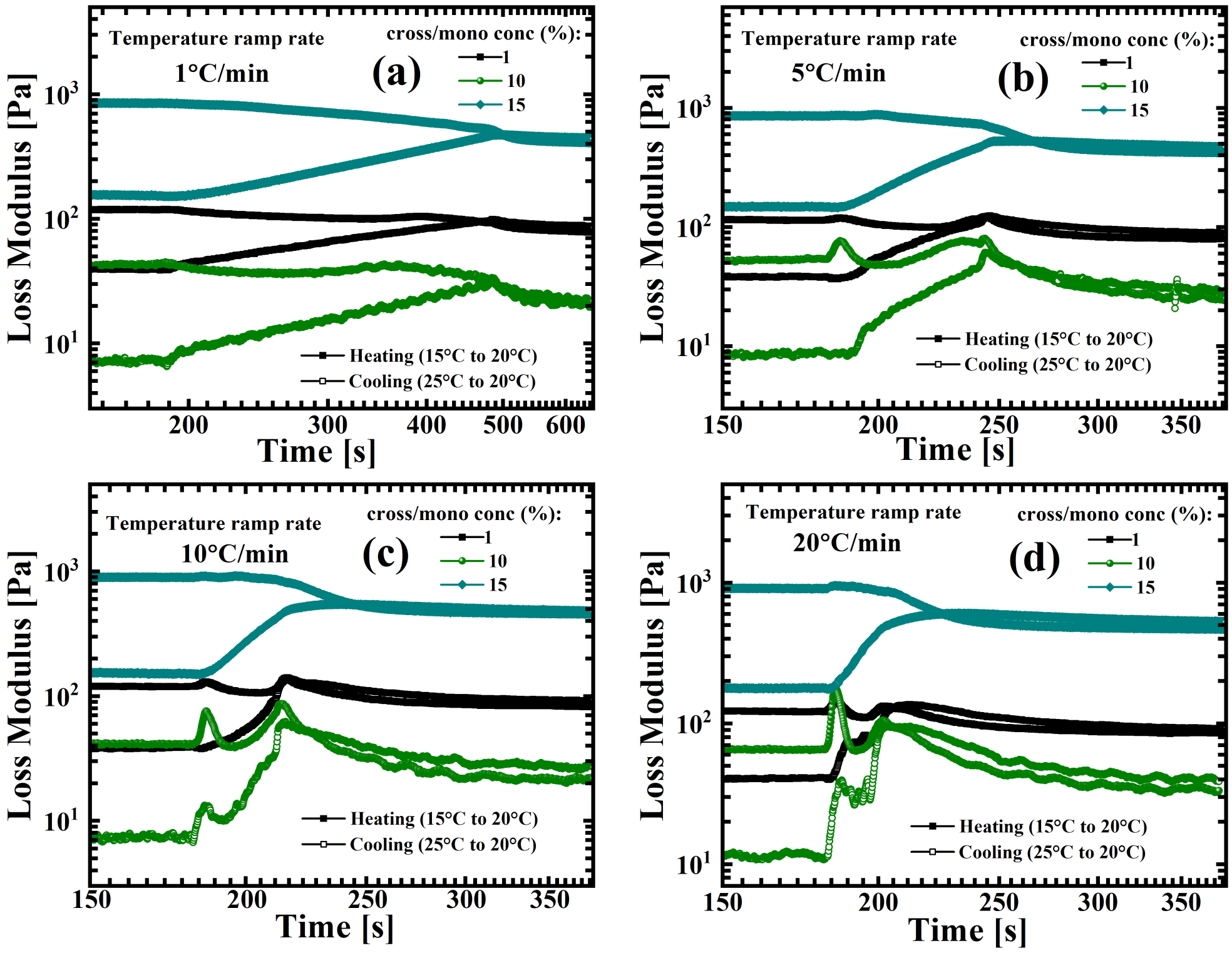


FIGURE S10: Loss modulus, G″, as a function of time during heating and cooling temperature ramps at rates: (a) 1°C/min (b) 5°C/min (c) 10°C/min and (d) 20°C/min, measured for dense PNIPAM suspensions prepared with particles of different stiffnesses at constant number density.

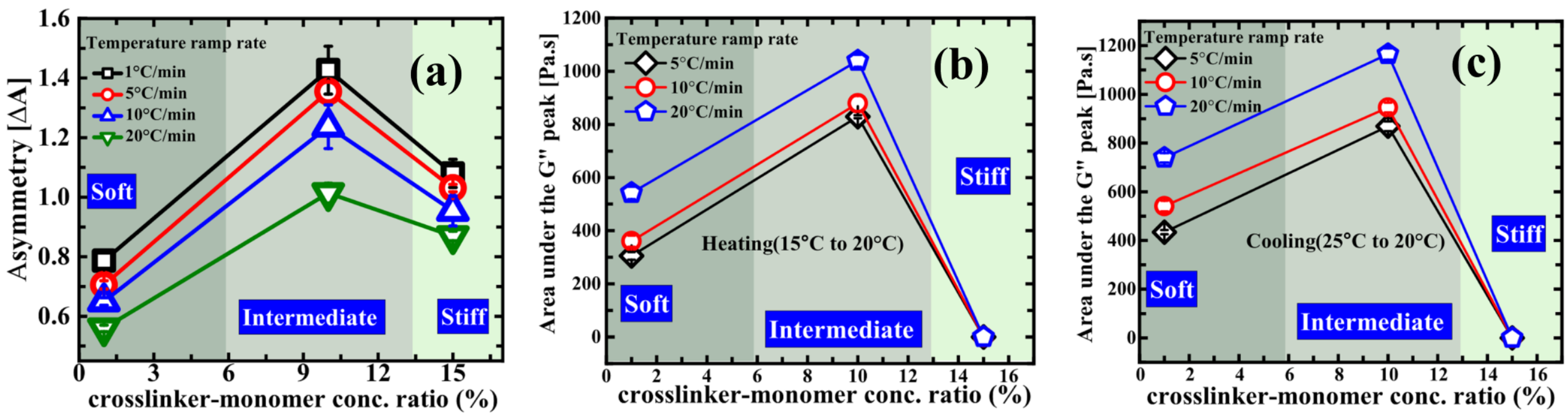


FIGURE S11: (a) Asymmetry of approach, ΔA, of storage modulus (G′), toward the final temperature $T_2$, under different applied heating and cooling ramp rates, for dense suspensions of PNIPAM particles of different stiffnesses prepared at constant particle number density. The areas under the loss modulus (G″-t) peaks during (b) heating and (c) cooling ramps, acquired simultaneously with the data presented in (a), as a function of microgel stiffness. The soft, intermediate, and stiff regimes are represented by distinct shades of green. The error bars denote standard deviations estimated from three independent measurements.

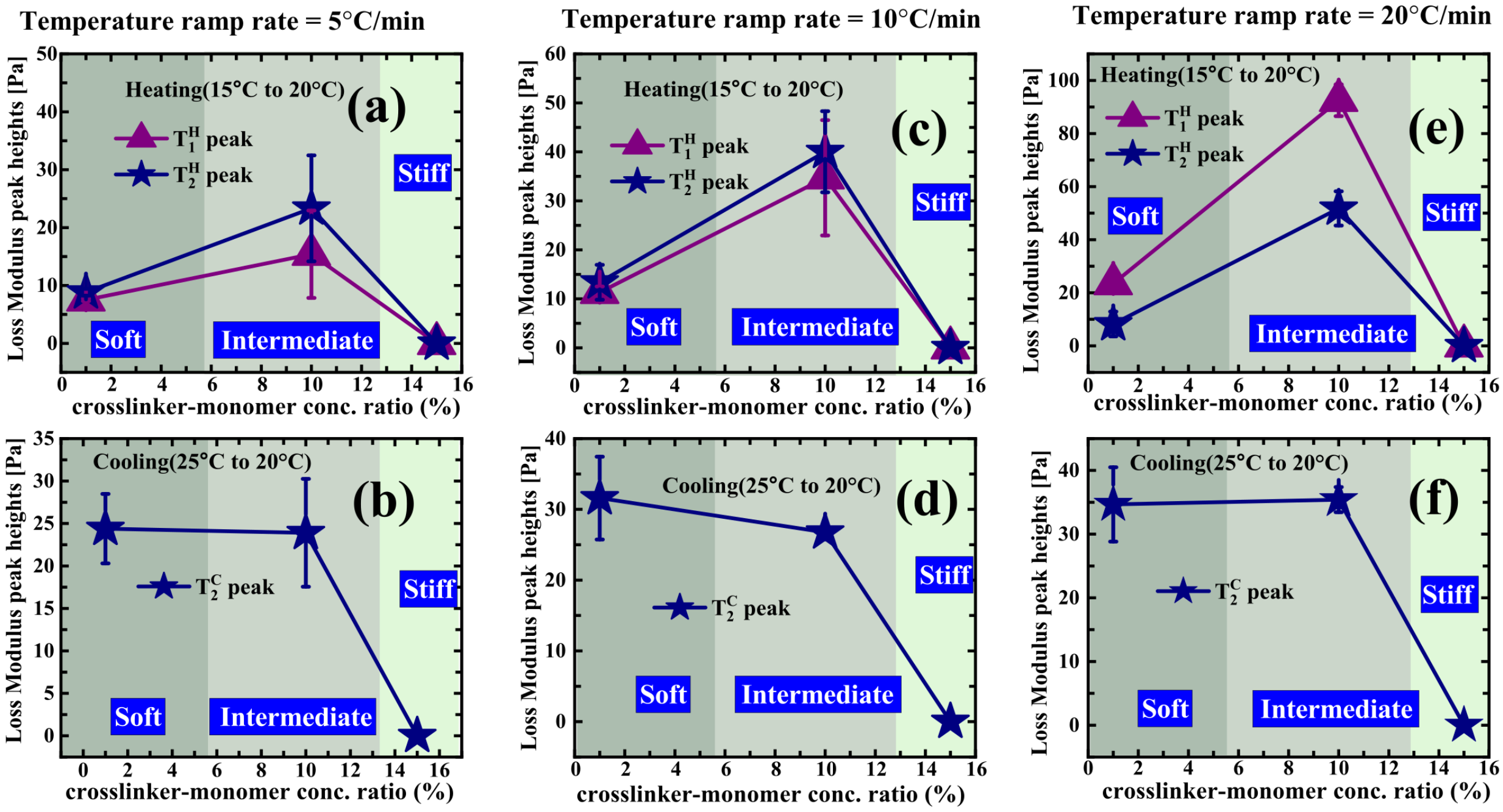


FIGURE S12: Loss modulus, G″, peak heights, measured across ramp rates, for dense PNIPAM suspensions comprising particles of different stiffnesses and prepared at constant particle number density. (a, b) Heating and cooling ramps at 5°C/min. (c, d) Heating and cooling ramps at 10°C/min. (e, f) Heating and cooling ramps at 20°C/min. The soft, intermediate, and stiff regimes are represented by distinct shades of green. Error bars represent standard deviations calculated from three independent measurements.

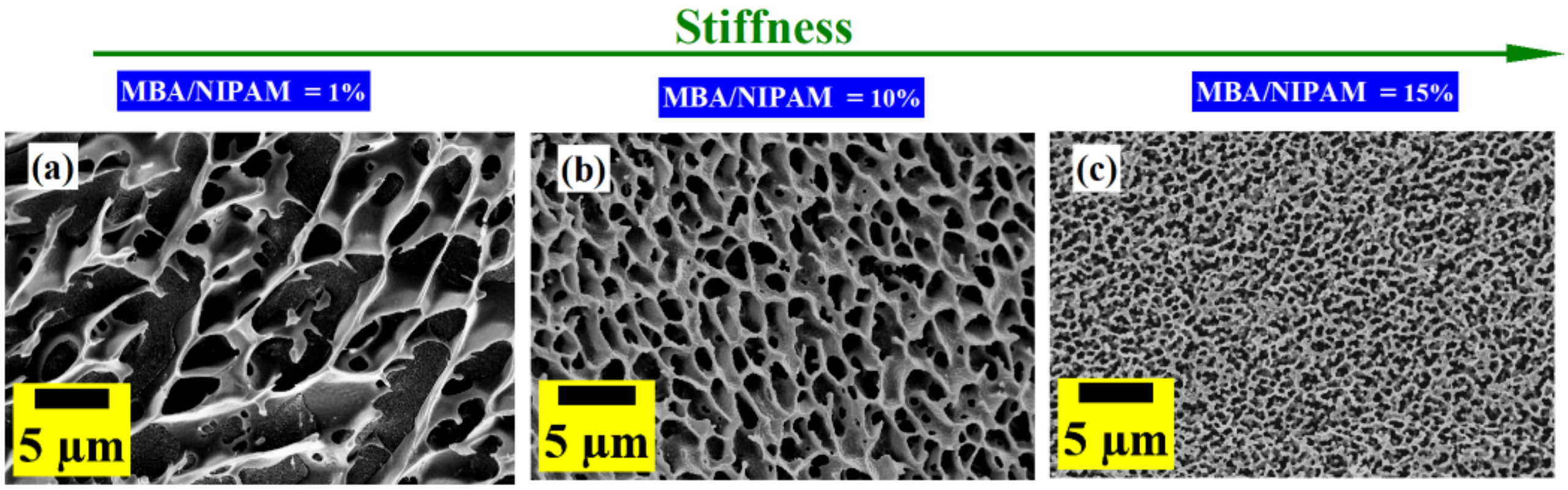


FIGURE S13: Cryo-FESEM micrographs of dense aqueous suspensions of PNIPAM microgel particles, prepared at a constant number density and acquired at 25°C with particle stiffnesses: (a) 1%, (b) 10%, and (c) 15% MBA/NIPAM, respectively. The green arrow points towards the increasing particle stiffness.


\end{document}